\documentclass[aps,showpacs,preprintnumbers,nofootinbibt,preprint]{revtex4}
\usepackage{amsmath,multirow,tabularx}
\usepackage{eurosym}
\usepackage{amssymb}
\usepackage{graphicx,epstopdf}
\usepackage{color}

\def\be{\begin{equation}}
\def\ee{\end{equation}}
\def\bea{\begin{eqnarray}}
\def\eea{\end{eqnarray}}

\begin{document}

\title{Testing the Bose-Einstein Condensate dark matter model at galactic cluster scale}
\author{Tiberiu Harko$^1$}
\email{t.harko@ucl.ac.uk}
\author{Pengxiang Liang$^2$}
\email{lpengx@mail2.sysu.edu.cn2}
\author{Shi-Dong Liang$^2$}
\email{stslsd@mail.sysu.edu.cn}
\author{Gabriela Mocanu$^3$}
\email{gabriela.mocanu@ubbcluj.ro}
\affiliation{$^1$ Department of Mathematics, University College London, Gower Street,
London, WC1E 6BT, United Kingdom}
\affiliation{$^2$State Key Laboratory of Optoelectronic Material and Technology, and
Guangdong Province Key Laboratory of Display Material and Technology, School
of Physics and Engineering,\\
Sun Yat-Sen University, Guangzhou 510275, People's Republic of China}
\affiliation{$^{3}$ Romanian Academy, Astronomical Institute, Astronomical Observatory Cluj-Napoca, 15 Cire\c{s}ilor Street, 400487 Cluj-Napoca,  Romania}
\date{\today }

\begin{abstract}
The possibility that dark matter may be in the form of a Bose-Einstein Condensate (BEC) has been extensively explored at galactic scale. In particular, good fits for the galactic rotations curves have been obtained, and upper limits for the dark matter particle mass and scattering length have been estimated. In the present paper we extend the investigation of the properties of the BEC dark matter to the galactic cluster scale, involving dark matter dominated astrophysical systems formed of thousands of galaxies each.  By considering that one of the major components of a galactic cluster, the intra-cluster hot gas, is described by King's $\beta$-model, and that both intra-cluster gas and dark matter are in hydrostatic equilibrium, bound by the same total mass profile, we derive the mass and density profiles of the BEC dark matter. In our analysis we consider several theoretical models, corresponding to isothermal hot gas and zero temperature BEC dark matter, non-isothermal gas and zero temperature dark matter, and isothermal gas and finite temperature BEC, respectively. The properties of the finite temperature BEC dark matter cluster are investigated in detail numerically. We compare our theoretical results with the observational data of 106 galactic clusters. Using a least-squares fitting, as well as the observational results for the dark matter self-interaction cross section,  we obtain some upper bounds for the mass and scattering length of the dark matter particle. Our results suggest that the mass of the dark matter particle is of the order of $\mu $eV, while the scattering length has values in the range of $10^{-7}$ fm.
\end{abstract}

\maketitle

\section{Introduction}

Dark matter, first introduced to explain the flat rotation curves at
galactic scales, and the virial mass discrepancy at the galaxy cluster
level, is presently considered to be a major component of the Universe.
Dark matter, assumed to be non-baryonic and non-relativistic, is  detected only by
its gravitational effects at the scale of galaxies and clusters
of galaxies, through observations of the motion of massive test particles \cite{d0}. However,
the particle nature of dark matter still remains mysterious. The most
popular candidates for the dark matter particles are the WIMP's, weakly
interacting massive, lying beyond the standard model of the particle physics \cite{d1}. Other
possible explanations for dark matter include self-interacting fermionic
dark matter \cite{d2}, scalar-field dark matter \cite{d3}, or modified
gravitational theories \cite{d4}.

In the standard $\Lambda $CDM ($\Lambda $ Cold Dark Matter) cosmological
scenario dark matter is assumed to be a pressureless, low temperature cold fluid,
interacting only gravitationally with itself, and with the cosmic
environment. However, as was shown in \cite{d5}, from the dark matter's lack
of deceleration in the Bullet Cluster collision, the self-interaction
cross-section $\sigma_{DM}$ of the dark matter particles with mass $m$ can be
constrained as $\sigma_{DM}/m < 1.25$ cm$^2$/g (68\% confidence limit) for
long-ranged forces. Moreover, by using the Chandra and Hubble Space
Telescopes observations of 72 collisions, a significant high dark matter
self-interaction cross-section $\sigma_{DM}/m < 0.47$ cm$^2$/g (95\% CL) was
obtained. From a theoretical point of view Self-Interacting Dark Matter
(SIDM) models were introduced in \cite{sidm}. An assumption which was
intensively investigated is the possibility that the dark matter
self-interaction is velocity dependent, with the simplest models assuming
cross sections of the form $1/v^{\alpha}$, where $v$ is the dark matter
velocity, and $\alpha $ is a constant \cite{d6}. N-body simulations of a
class of self-interacting dark matter models, having a non-power-law velocity
dependence of the transfer cross section, physically motivated by a
Yukawa-like gauge boson interaction,  were performed in \cite{d6}. More
exactly, and contrary to the standard cold dark matter model case, for
self-interacting dark matter there are no subhalos that are more
concentrated, as compared to what is inferred from the kinematics of the
Milky Way Dwarf Spheroidal Galaxies. The rate of dark matter scattering in collapsed structures
throughout the history of the Universe was estimated recently in \cite{d7}. If the
scattering cross-section is velocity-independent, it turns out that dark
matter particle scattering occur mainly at late times in the cosmological
evolution of the Universe. On the other hand, for dark matter models with a
velocity-dependent cross-section (like, for example, the Yukawa potential
interactions via a massive mediator), the scattering rate has a maximum at
around $z\sim 20$.  This maximum happens in objects with mass less than $10^4
M_{\odot}$.

However, from both a fundamental theoretical point of view, and from a
phenomenological perspective, \textit{the physically best motivated
self-interacting dark matter model is represented by the Bose-Einstein
Condensate dark matter model}.

The idea of the Bose-Einstein Condensation was first introduced from a statistical
physical point of view in the 1920s \cite{Bose,Ein}. The experimental
realization of the Bose-Einstein Condensation in dilute alkali gases was
achieved seventy years later, in 1995 \cite{exp}. A very basic and general result in quantum
statistical physics is that at very low temperatures, in a dilute Bose gas,
all bosonic particles condense to the same quantum ground state, forming a
Bose-Einstein Condensate (BEC). From a physical point of view the realization of a  BEC is
signalled  by sharp peaks in both coordinate and momentum space
distributions. The Bose-Einstein Condensation process starts when particles
become correlated quantum mechanically,  that is, when their de Broglie thermal wavelength is
greater than the mean particle distance. Equivalently, the condensation is
initiated when the temperature $T$ of the system is lower than the transition one, $T_{tr}$, given by
\begin{equation}  \label{Ttr}
T_{tr}=\frac{2\pi\hbar^2\rho_{tr}^{2/3}}{ \zeta^{2/3}(3/2)m^{5/3}k_B},
\end{equation}
where $m$ is the particle mass in the condensate, $\rho_{tr}$ is the
transition density, $k_B$ is Boltzmann's constant, and $\zeta $ denotes the
Riemmann zeta function.

Generally, and up to now, BEC was observed and studied extensively in the laboratory, on a very
small scale, rather than at a large galactic/cosmological scale. However, \textit{the
possibility that bosons also condensate on cosmological or astrophysical
scales cannot be rejected a priori}. Therefore, based on the present day
theoretical and experimental knowledge of the Bose-Einstein condensation
processes, it is reasonable to assume the presence of some forms of
condensates in the cosmic environment. Due to their superfluid properties
some stellar type astrophysical objects, like neutron or quark stars, may
contain a significant part of their mass in the form of a Bose-Einstein
Condensate. Stars made of Bose-Einstein Condensates were investigated in
\cite{stars}. Condensate stars with particle masses of the order of two
neutron masses, forming Cooper pairs, and with scattering length of the
order of 10-20 fm may have maximum masses of the order of 2 $M_{\odot}$,
maximum central densities of the order of $0.1-0.3\times 10^{16}$ g/cm$^3$,
and minimum radii in the range of 10-20 km.

The idea that dark matter is in the form of a Bose-Einstein Condensate was
considered, mostly from a phenomenological point of view, in \cite{early}. A
systematic study of the properties of the BEC dark matter was initiated in
\cite{BoHa07}. In order to study the dark matter condensate the
non-relativistic Gross-Pitaevskii (GP) in the presence of a
confining gravitational potential  was considered  as a starting point. A significant simplification of the
mathematical formalism can be achieved by introducing the Madelung
representation of the wave function, a standard approach in condensed matter physics. Then the GP equation can be represented
in the equivalent form of a continuity equation, and of a hydrodynamic Euler type
equation. From the Madelung representation of the GP equation it follows that dark matter
can be described as a non-relativistic, Newtonian Bose-Einstein
gravitational condensate gas, whose density and pressure are related by a
barotropic equation of state. To test the validity of the BEC dark
matter model at the galactic scales the Newtonian tangential velocity
equation was fitted with a sample of rotation curves of low surface
brightness and dwarf galaxies, respectively. A good agreement was found
between the theoretical rotation curves, and the observational data \cite%
{BoHa07}.

The properties of BEC dark matter were further investigated in \cite{inv,
Har1}. In particular, in \cite{Har1} it was shown that the non-singular
density profiles of the Bose-Einstein condensed dark matter generally show
the presence of an extended core, whose formation is explained by the strong
interaction between dark matter particles. The mean value of the logarithmic
inner slope of the mass density profile of dwarf galaxies was also obtained,
and it was shown that the observed value of this parameter is in agreement
with the theoretical estimations obtained in the framework of the BEC dark
matter model. The cosmological implications of the finite temperature BEC
dark matter were considered in \cite{Har2}.

An interesting result in the theory of the zero temperature BEC
dark matter is that in the case of a condensate with quartic non-linearity,
the equation of state is polytropic with index $n=1$ \cite{BoHa07}. Therefore, the
corresponding Lane-Emden equation, describing the gravitational properties
of the condensate, can be solved exactly, and the density profile of the
dark matter can be obtained in a simple form. Once the BEC dark matter
density profile is known, all the physical parameters of the condensed
system (mass, radius, central density) as well as the rotational speeds can
be obtained, thus leading to the possibility of a full observational test of
the model.

Up to now the properties of the BEC dark matter were investigated mostly on
the galactic scale. However, the second important evidence for dark matter
comes from the virial mass discrepancy in galaxy clusters \cite{BT}. A
galaxy cluster is a giant astrophysical object formed of hundreds to thousands
of galaxies, bounded together by their own gravitational interaction. They
are composed of galaxies, representing around 1\% of their mass, high
temperature intracluster gas, representing around 9\% of the cluster mass,
and dark matter, representing 90\% of the their mass. The total masses
obtained by measuring the velocity dispersions of the galaxies exceed the
total masses of all stars in the cluster by factors of order of $\sim$ 200 -
400 \cite{revc}. The measurement of the temperature of the intracluster
medium represents another strong evidence for the presence of dark matter,
since the determined depth of the gravitational potential of the clusters
requires a supplementary mass component \cite{revc}. Hence galaxy clusters,
being dark matter dominated astrophysical structures, are the ideal testing ground for the
properties of dark matter.

It is the purpose of the present paper to consider the properties of the BEC
dark matter at the galactic cluster scale. Thus we extend the previous investigations of the BEC dark matter properties at the galactic scale to astrophysical systems consisting of hundreds to thousands galaxies. We begin our study with a brief review of the properties of self-interacting dark matter bosonic system in BEC state. From the non-relativistic Gross-Pitaevskii
equation and its Madelung representation of the wave function, we obtain the
pressure of the system for both zero temperature and finite temperature BEC dark matter. We assume that the BEC dark matter systems are bound
to hydrostatic equilibrium by the gravitational potential of the cluster. Meanwhile,
we consider that the hot intracluster gas, with density profile described by the standard King's $\beta$-model, is bound to equilibrium in the same gravitational potential. Thus, the knowledge of the hot gas distribution allows as to reconstruct the density and mass distribution of the BEC dark matter in the cluster. This reconstruction is model dependent, and in the present paper we consider three distinct cases: the case of the isothermal gas in gravitational equilibrium with zero temperature BEC dark matter, non-isothermal gas in equilibrium with zero temperature BEC dark matter, and finally, the case of the isothermal gas in equilibrium with finite temperature BEC dark matter. In each of these cases the dark matter density and mass profiles are obtained, with the radius of the cluster  expressed as a function of the astrophysical parameters of the cluster. The properties of the finite temperature BEC dark matter halos in galactic clusters are investigated in detail by using numerical methods. The  fitting of the model predictions  via the least-$\chi^2$ fitting with the observational data for 106 clusters \cite{RB} allows the direct determination of the ratio $\lambda =m^3/a$ only, where $m$ is the mass of the dark matter particle, and $a$ is the scattering length. By combining the obtained values of $\lambda $ with the observational self interaction cross sections, we can obtain some upper limits of the mass and scattering length of the dark matter particles.

The present paper is organized as follows. In Section~\ref{sect1} we briefly review the basic physical and astrophysical properties of the BEC dark matter, and we present the relevant equations of state for both zero temperature and finite temperatures cases. The zero temperature dark matter properties are investigated in Section~\ref{sect2} for the isothermal  intracluster gas case. The cluster dark matter density and mass profiles are obtained analytically, and the theoretical predictions are compared with the observations. The effects of the variation of the gas temperature on the astrophysical parameters of the zero temperature BEC dark matter are investigated in Section~\ref{sectn}. The case of the finite temperature BEC dark matter is considered in Section~\ref{sect3}. We discuss and conclude our results in Section~\ref{sect4}.

\section{Basic properties Bose-Einstein Condensate dark matter at zero and
finite temperature}\label{sect1}

In the present Section we briefly review the main properties of the
Bose-Einstein Condensate dark matter at both zero and finite temperature, and we present the results
which will be used in the sequel. For an in-depth discussion of the
considered issues, and of the detailed derivation of the main results, we
further refer the reader to the papers and books \cite{Gr}-{\cite{HFB}. }

\subsection{The generalized Gross-Pitaevskii equation and the hydrodynamic
representation}

The starting point in the study of the Bose-Einstein Condensates is the Heisenberg
equation of motion for the quantum field operator $\hat{\Phi}\left(\vec{r}%
,t\right)$. At arbitrary temperatures the Heisenberg equation, describing the
dynamics of a Bose-Einstein condensate, is given by \cite{Za99, HaM, Prouk,
Gr}
\begin{eqnarray}
i\hbar \frac{\partial \hat{\Phi}\left(\vec{r},t\right) }{\partial t}&=&%
\Bigg[ -\frac{\hbar ^{2}}{2m}\Delta +mV_{grav}\left(\vec{r},t\right) +
g^{\prime }\hat{\Phi}^{+}\left( \vec{r},t\right) \hat{\Phi}\left(\vec{r}%
,t\right) \Bigg] \hat{\Phi}\left( \vec{r},t\right) ,  \label{1n}
\end{eqnarray}
where $m$ is the mass of the particle in the condensate, $V_{grav}\left(
\vec{r},t\right) $ is the external gravitational trapping potential, and $%
g^{\prime }=4\pi a\hbar ^{2}/m$. In the following we denote by $a$ the $s$%
-wave scattering length. In order to obtain Eq.~(\ref{1n}) we have assumed
that the inter-particle interaction potential is represented as a zero-range
pseudo-potential of strength $g^{\prime }$. As a next step in our study we take the
average of Eq.~(\ref{1n}) with respect to a non-equilibrium, broken symmetry
ensemble. Over this ensemble the quantum field operator takes a non-zero
expectation value. We denote the average of the field operator $\hat{\Phi}%
\left(\vec{r},t\right)$ by $\Psi \left(\vec{r},t\right)$, $\Psi \left(\vec{r}%
,t\right) =\left\langle \hat{\Phi}\left( \vec{r},t\right) \right\rangle $.
From a physical point of view $\Psi \left(\vec{r},t\right)$ represents the
condensate wave-function. After averaging Eq.~(\ref{1n}) we find the exact equation of motion
of $\Psi \left( \vec{r},t\right)$  as
\begin{eqnarray}
i\hbar \frac{\partial \Psi \left( \vec{r},t\right) }{\partial t}&=&\left[ -%
\frac{\hbar ^{2}}{2m}\Delta +mV_{grav}\left( \vec{r},t\right) \right] \Psi
\left( \vec{r},t\right) + g^{\prime }\left\langle \hat{\Phi}^{+}\left( \vec{r%
},t \right) \hat{\Phi}\left( \vec{r},t\right) \hat{\Phi}\left( \vec{r}%
,t\right) \right\rangle .  \label{2a}
\end{eqnarray}

Now we introduce the non-condensate field operator $\tilde{\psi}\left( \vec{r%
},t \right) $ by means of the definition $\hat{\Phi}\left( \vec{r},t\right)
=\Psi \left(\vec{r},t\right) +\tilde{\psi}\left( \vec{r},t\right)$.
Furthermore we assume that the average value of $\tilde{\psi}\left( \vec{r}%
,t\right) $ is zero, $\left\langle \tilde{\psi}\left(\vec{r},t\right)
\right\rangle =0$. Therefore we can separate out the condensate component of
the quantum field operator, and thus we obtain the equation of motion for $%
\Psi $ in the form \cite{Za99,HaM,Prouk, Gr}
\begin{eqnarray}
i\hbar \frac{\partial \Psi \left(\vec{r},t\right) }{\partial t}&=&\Bigg[ -%
\frac{\hbar ^{2}}{2m}\Delta +mV_{grav}\left( \vec{r},t\right) +g\rho
_{c}\left( \vec{r},t\right) + 2g\tilde{\rho}\left( \vec{r},t\right) \Bigg] %
\Psi \left( \vec{r},t\right) +  \notag \\
&&g\rho _{\tilde{m}}\left( \vec{r},t\right)\Psi ^{\ast }\left( \vec{r}%
,t\right)+ g\rho _{\tilde{\psi}^{+} \tilde{\psi} \tilde{\psi}}\left( \vec{r}%
,t\right) ,  \label{6n}
\end{eqnarray}
where we have denoted $g=4\pi a\hbar ^{2}/m^{2}$. In Eq.~(\ref{6n}) we have also
introduced the four densities characterizing the condensate \cite{Za99,HaM,Prouk, Gr}, namely, the local
condensate mass density
\begin{equation}
\rho _{c}\left( \vec{r},t\right) =mn_{c}\left( \vec{r},t\right) =m\left|
\Psi \left( \vec{r},t\right) \right| ^{2},
\end{equation}
the non-condensate mass density
\begin{equation}
\tilde{\rho}\left( \vec{r},t\right) =m\tilde{n}\left(\vec{r},t\right)
=m\left\langle \tilde{\psi}^{+}\left( \vec{r},t\right) \tilde{\psi}\left(
\vec{r},t\right) \right\rangle ,
\end{equation}
the off-diagonal (anomalous) mass density
\begin{equation}
\rho _{\tilde{m}}\left( \vec{r},t\right)=m\tilde{m}\left( \vec{r},t\right)
=m\left\langle \tilde{\psi}\left( \vec{r},t\right) \tilde{\psi}\left( \vec{r}%
,t\right) \right\rangle ,
\end{equation}
and, finally, the three-field correlation function density
\begin{equation}
\rho _{\tilde{\psi}^{+} \tilde{\psi} \tilde{\psi}}\left( \vec{r}%
,t\right)=m\left\langle \tilde{\psi}^{+}\left( \vec{r},t\right) \tilde{\psi}%
\left( \vec{r},t\right) \tilde{\psi}\left( \vec{r},t\right) \right\rangle,
\end{equation}
respectively.

In the present paper we will restrict our analysis to the range of finite temperatures
where the dominant thermal excitations can be approximated as non-condensed
particles of high energy, with the particles evolving in a self-consistent exterior
Hartree-Fock mean field, whose local energy is \cite{Za99, HaM, Prouk, Gr}
\begin{eqnarray}
\bar{\varepsilon}_{p}\left(\vec{r},t\right) &=&\frac{\vec{p}^{2}}{2m}%
+mV_{grav}\left( \vec{r},t\right) +2g\left[ \rho _{c}\left( \vec{r},t\right)
+\tilde{\rho}\left( \vec{r},t\right) \right] = \frac{\vec{p}^{2}}{2m}%
+U_{eff}\left( \vec{r},t\right) ,
\end{eqnarray}
where $U_{eff}\left( \vec{r},t\right) =mV_{grav}\left( \vec{r},t\right) +2g%
\left[ \rho _{c}\left( \vec{r},t\right) +\tilde{\rho}\left( \vec{r},t\right) %
\right] $. Therefore, in the present analysis we neglect the Hartree-Fock
type mean field effects associated to the anomalous density $\rho _{\tilde{m}%
}$, and with the three-field correlation function $\left\langle \tilde{\psi}%
^{+}\tilde{\psi}\tilde{\psi}\right\rangle $, respectively. From the point of
view of astrophysical applications, for dark matter halos having a large
number of particles this represents a very good approximation, since one can
show that the contribution of the anomalous density and of the three-field
correlation function to the total density is just of the order of a few
percents \cite{HaM}.

\subsubsection{Equilibrium properties of finite temperature Bose-Einstein
Condensates}

One of the important properties of the finite temperature Bose-Einstein
Condensates is that in the thermal cloud the collision between particles
determine the non-equilibrium distribution to evolve towards the standard
static Bose-Einstein distribution function $f^{0}\left( \vec{r},\vec{p}%
\right) $ \cite{Gr}. This means that the particles in the thermal cloud are
in thermodynamic equilibrium between themselves. With the use of a
single-particle representation spectrum it follows that the equilibrium
distribution of the thermal cloud can be represented as \cite{Za99,Gr}
\begin{equation}
f^{0}\left(\vec{p}, \vec{r},t\right) =\left[ e^{\beta \bar{\varepsilon}%
_{p}\left( \vec{r},t\right) -\tilde{\mu}}-1\right] ^{-1},
\end{equation}
where $\beta =1/k_{B}T$, and $\tilde{\mu}$ is the chemical potential of the
thermal cloud. In order to determine the chemical potential $\tilde{\mu}$ we
make the simplifying physical assumption that the condensate and all the thermal
cloud components are in local diffusive equilibrium with respect to each
other. Thus, the requirement of a diffusive equilibrium between the cloud
and the condensate gives the thermodynamic condition $\mu _{c}=\tilde{\mu}$,
where $\mu _c$ is the chemical potential of the condensate. Therefore it
follows that the chemical potential $\mu _c$ of the condensate also
determines the static equilibrium distribution of the particles in the
thermal cloud \cite{Za99,HaM,Prouk, Gr}.

The equilibrium density of the thermal excitations is obtained from the
equilibrium Bose - Einstein distribution by integration over the momenta.
Therefore we obtain \cite{Gr, HaM, Za99,Prouk}
\begin{equation}
\tilde{\rho}\left( \vec{r},t\right) =\frac{m}{\left(2\pi \hbar \right)^{3}}%
\int d^{3}\vec{p}f^{0}\left( \vec{p}, \vec{r},t\right) =\frac{m}{\lambda
_T^{3}}g_{3/2}\left[ z\left( \vec{r},t\right) \right] ,
\end{equation}
where $\lambda _T=$ $\sqrt{2\pi \hbar ^{2}\beta /m}$ is the de Broglie
thermal wavelength, $g_{3/2}(z)$ is a so-called Bose-Einstein function, and
the fugacity $z\left( \vec{r},t\right) $ is defined as
\begin{equation}
z\left(\vec{r},t\right) =e^{\beta \left[ \tilde{\mu}-U_{eff}\left( \vec{r}%
,t\right) \right] }=e^{-\beta g\rho _{c}\left( \vec{r},t\right) }.
\end{equation}

The pressure $\tilde p$ of the thermal excitations of the BEC dark matter can be computed from the
standard statistical physics definition \cite{Gr,Za99}
\begin{equation}
\tilde{p}\left( \vec{r},t\right) =\int \frac{d\vec{p}}{\left( 2\pi \hbar
\right) ^{3}}\frac{p^{2}}{3m}f^{0}\left( \vec{p},\vec{r},t\right) ,
\end{equation}
and is given by
\begin{equation}
\tilde{p}\left( \vec{r},t\right) =\frac{1}{\beta \lambda _{T}^{3}}g_{5/2}%
\left[ z\left( \vec{r},t\right)\right] .
\end{equation}

The Bose-Einstein functions $g_{3/2}\left(e^{-x}\right)$ and $%
g_{5/2}\left(e^{-x}\right)$ can be easily computed numerically by using the following
series expansions \cite{Rob},
\begin{eqnarray}
g_{3/2}\left( e^{-x}\right) =2.612-3.544\sqrt{x}+1.460x-
0.103x^{2}+0.00424x^{3}+O\left(x^{7/2}\right) .  \label{expbe1}
\end{eqnarray}
and
\begin{eqnarray}  \label{expbe2}
g_{5/2}\left( e^{-x}\right) =1.341+2.363x^{3/2}-2.612x-
0.730x^{2}+0.0346x^{3}+O\left( x^{7/2}\right) ,
\end{eqnarray}
respectively. For $x<1$, Eqs.~(\ref{expbe1}) and (\ref{expbe2}) approximates
the function $g_{3/2}\left( e^{-x}\right) $ and $g_{5/2}\left(e^{-x}\right) $
with an error smaller than 1\%.

\subsubsection{The hydrodynamic representation for finite temperature
Bose-Einstein Condensates}

The Gross-Pitaevskii equation for finite temperature condensates can be
transformed into an equivalent hydrodynamic form with the help of the
Madelung representation of the wave function given by $\Psi \left( t,\vec{r}%
\right) =\sqrt{\rho _{c}}\exp \left[ \left( i/\hbar \right) S\left( \vec{r}%
,t\right) \right] $. Then, by neglecting the effects of the mean field
associated with the anomalous density and the three-field correlation
function, respectively, it follows that Eq.~(\ref{6n}) can be represented as
a hydrodynamic type system, given by \cite{Za99, HaM, Gr}
\begin{equation}
\frac{\partial \rho _{c}}{\partial t}+\nabla \cdot \left( \rho _{c}\vec{v}%
_{c}\right) =0,
\end{equation}
\begin{equation}
\frac{\partial S}{\partial t}=-\left( \mu _{c}+\frac{1}{2}m\vec{v}%
_{c}^{2}\right) ,  \label{7n}
\end{equation}
where the local velocity of the condensate is defined as $\vec{v}_{c}\left(%
\vec{r},t\right) =\left( \hbar /m\right) \nabla S$. The chemical potential
of the condensate is obtained from the relation
\begin{equation}
\mu _{c}=-\frac{\hbar ^{2}}{2m}\frac{\Delta \sqrt{\rho _{c}}}{\sqrt{\rho _{c}%
}}+mV_{grav}\left( \vec{r},t\right) +g\rho _{c}\left( \vec{r},t\right) +2g%
\tilde{\rho}\left(\vec{r},t\right) .
\end{equation}

Eq.~(\ref{7n}) can be reformulated as the Euler equation of fluid dynamics
for the condensate,
\begin{equation}  \label{euler}
m\frac{d\vec{v}_{c}}{dt}=m\left[ \frac{\partial \vec{v}_{c}}{\partial t}%
+\left( \vec{v}_{c}\cdot \nabla \right) \vec{v}_{c}\right] =-\nabla \mu _{c}.
\end{equation}

The gravitational potential $V_{grav}$ satisfies the Poisson equation,
\begin{equation}  \label{pois}
m\Delta V_{grav}=4\pi G\left(\rho _c+\tilde{\rho}\right).
\end{equation}

The equation of state of the finite temperature condensate can be obtained from the hydrodynamical representation by using the Thomas-Fermi approximation for the condensate wave function \cite{Gr} - \cite{HFB}. In this approximation, the
kinetic energy term $-\left( \hbar ^{2}/2m\right) \Delta $ of the condensate
particles is considered as being negligibly small. Hence for the chemical potential of the condensate we find \cite{HaM}
\begin{equation}\label{mu}
\mu _{c}=mV_{grav}\left( \vec{r},t\right) +g\rho _{c}\left( \vec{r},t\right) +2g\tilde{%
\rho}\left(\vec{r},t\right),
\end{equation}

Therefore in the Thomas-Fermi approximation the Euler equation Eq.~(\ref{euler})  takes the form
\begin{equation}\label{euler1}
\rho_c\frac{d\vec{v}_{c}}{dt}=-\frac{g}{m}\rho_c\nabla\left[\rho _{c} +2\tilde{%
\rho}\right]-\rho _c\nabla V _{grav}.
\end{equation}

With the use of Eq.~(\ref{mu}), and by taking into account the explicit expression of $\tilde{\rho}$, it follows that the equation of motion of the Bose-Einstein condensate, given by Eq.~(\ref{euler1}),  can be written as
\be
\rho_c\frac{d\vec{v}_{c}}{dt}=-\nabla p_c-\rho _c\nabla V _{grav},
\ee
where we have introduced the pressure $p_c$ of the finite temperature Bose-Einstein condensate in thermal equilibrium with a gas of thermal excitations, defined as \cite{HaM}
\bea\label{pc}
p_c\left(T,\rho _c\right)&=&\frac{g}{2m}\rho _c^2-2.362\frac{g^{3/2}}{\lambda _T^3}\left(k_BT\right)^{-1/2}\rho _c^{3/2}+ 1.460\frac{g^2}{\lambda _T^3}\left(k_BT\right)^{-1}\rho _c^2-0.137\frac{g^3}{\lambda _T^3}\times \nonumber\\ &&\left(k_BT\right)^{-2}\rho _c^3+
0.00636\frac{g^4}{\lambda _T^3}\left(k_BT\right)^{-3}\rho _c^4.
\eea

\subsection{Physical properties of the dark matter condensate particle}

The simplest Bose-Einstein Condensate dark matter model can be obtained by
adopting the zero-temperature approximation. This approximation gives a
good description of the rotation curves of the galactic dark matter halos \cite{BoHa07, Har1}. It
also allows us, by using the galactic global astrophysical parameters (mass and radius),
to make an estimate of the physical properties of the dark matter particle.
By assuming that $\tilde{\rho}\equiv 0$, the condensate is static ($\vec{v}%
_c=0$, spherically symmetric, and by neglecting the quantum pressure term $%
\left(\hbar ^{2}/2m\right)\left(\Delta \sqrt{\rho _{c}}/\sqrt{\rho _{c}}%
\right)$, from Eqs.~(\ref{euler}) and (\ref{pois}) it follows that the
condensate dark halo density satisfies the Lane-Emden type differential equation \cite{BoHa07}
\begin{equation}  \label{rhoc}
\frac{1}{r^2}\frac{d}{dr}\left(r^2\frac{d\rho _c}{dr}\right)+\frac{Gm^3}{%
a\hbar ^2}\rho _c=0.
\end{equation}
From the analysis of the properties of the static Bose-Einstein condensate
dark matter halos as described by the solutions of Eq.~(\ref{rhoc}) it
follows that the radius $R$ of the zero temperature condensate dark matter
halo is given by $R=\pi \sqrt{\hbar ^{2}a/Gm^{3}}$ \cite{BoHa07}. The total
mass of the $T=0$ condensate dark matter halo $M$ is found as $M=4\pi
^2\left(\hbar ^2a/Gm^3\right)^{3/2}\rho _c=4R^3\rho _{c}(0)/\pi $, where $%
\rho _{c}(0)$ is the central density of the galactic halo. The mean value $%
<\rho >$ of the density of the zero temperature condensate is given by the
expression $<\rho >=3\rho _{c}(0)/\pi ^2$. From the previous results it
follows that the dark matter particle mass in the condensate satisfies a
mass-galactic radius relation of the form \cite{BoHa07}
\begin{eqnarray}  \label{mass}
m &=&\left( \frac{\pi ^{2}\hbar ^{2}a}{GR^{2}}\right) ^{1/3}\approx
6.73\times 10^{-2}\times \left[ a\left( \mathrm{fm}\right) \right] ^{1/3}%
\left[ R\;\mathrm{(kpc)}\right] ^{-2/3}\;\mathrm{eV}.
\end{eqnarray}
For $a\approx 1 $ fm and $R\approx 10$ kpc, we obtain a simple estimate of
the mass of the condensate particle as being of the order of $m\approx 14$
meV. On the other hand, for $a\approx 10^{6}$ fm, corresponding to the
values of $a$ observed in terrestrial laboratory experiments with rubidium
and cesium cold gases, we have $m\approx 1.44$ eV.

An important observational method that can be used very successfully to
obtain the physical properties of dark matter is the study of the collisions
between clusters of galaxies. Typical examples of such processes are the
Bullet Cluster (1E 0657-56) and the Baby Bullet (MACSJ0025-12),
respectively \cite{Bul, Bul1}. These astrophysical studies provide important constraints on
the physical properties of dark matter, like, for example, its interaction
cross-section with normal baryonic matter, as well as the dark matter-dark
matter self-interaction cross section. If from observations one can obtain
the ratio $\sigma _m=\sigma /m$ of the self-interaction cross section $%
\sigma =4\pi a^2$ and of the dark matter particle mass $m$, then the mass of
the dark matter particle in the Bose-Einstein Condensate galactic halo can
be estimated from Eq.~(\ref{mass}) as \cite{HaM}
\begin{equation}
m=\left(\frac{\pi ^{3/2}\hbar ^2}{2G}\frac{\sqrt{\sigma _m}}{R^2}%
\right)^{2/5}.
\end{equation}

By analyzing several sets of results from X-ray, strong lensing, weak
lensing, and optical observations with the complex numerical simulations of
the merging of the Bullet Cluster, an upper limit (68 \% confidence) for $%
\sigma _m$ of the order of $\sigma _m<1.25\;\mathrm{cm^2/g}$ was determined
in \cite{Bul}. By adopting for $\sigma _m$ a value of $\sigma _m=1.25\;%
\mathrm{cm^2/g}$, the mass of the dark matter particle in the cold
Bose-Einstein Condensate dark matter halo can be constrained as having an
upper limit of the order
\begin{eqnarray}\label{massg}
m&<&3.1933\times10^{-37}\left(\frac{R}{10\;\mathrm{kpc}}\right)^{-4/5}\times
\left(\frac{\sigma _m}{1.25\;\mathrm{cm^2/g}}\right)^{1/5}\;\mathrm{g}=
0.1791\times\left(\frac{R}{10\;\mathrm{kpc}}\right)^{-4/5} \times  \notag \\
&&\left(\frac{\sigma _m}{1.25\;\mathrm{cm^2/g}}\right)^{1/5}\;\mathrm{meV}.
\end{eqnarray}
By using the above value for the particle mass we can constrain the
scattering length $a$ as
\begin{equation}\label{ag}
a<\sqrt{\frac{\sigma _m\times m}{4\pi }}=1.7827\times 10^{-19}\;\mathrm{cm}%
=1.7827\times 10^{-6}\;\mathrm{fm}.
\end{equation}
Therefore it follows that the value of the scattering length $a$, obtained
from the astrophysical observations of the Bullet Cluster, is much smaller
than the value of $a=10^4-10^6$ fm, which characterises cold Bose-Einstein
Condensates in terrestrial laboratory experiments \cite{exp}.

A stronger constraint for the self-interaction cross section of the dark
matter particles $\sigma _m$ was proposed in \cite{Bul1}, with $\sigma
_m\in(0.00335\;\mathrm{cm^2/g},0.0559\;\mathrm{cm^2/g})$. This cross section
range gives a dark matter particle mass of the order
\begin{eqnarray}\label{massg1}
m&\approx &\left(9.516\times 10^{-38}-1.670\times 10^{-37}\right)\left(\frac{%
R}{10\;\mathrm{kpc}}\right)^{-4/5}\;\mathrm{g}=
\left(0.053-0.093\right)\left(\frac{R}{10\;\mathrm{kpc}}\right)^{-4/5}\;%
\mathrm{meV},  \notag \\
\end{eqnarray}
while the scattering length is of the order of
\begin{eqnarray}\label{ag1}
a&\approx &\left(5.038-27.255\right)\times 10^{-21}\;\mathrm{cm}%
=\left(5.038-27.255\right)\times 10^{-8}\;\mathrm{fm}.
\end{eqnarray}

\section{Zero temperature BEC dark matter in galactic clusters}\label{sect2}

A large number of astronomical and astrophysical observations have shown
that galaxies tend to concentrate in larger structures, called clusters of
galaxies, with total masses ranging from $10^{13}M_{\odot}$ for groups, and
up to a few $10^{15}M_{\odot}$ for very large systems. The morphology of a
galactic cluster is generally dominated by a regular, centrally peaked main
massive component \cite{cl1,RB}. Usually clusters are considered to be
dark matter dominated astrophysical systems. Therefore their formation and evolution is largely controlled by the
gravitational interaction between their mass components. The initial
conditions of the cluster mass distribution are already set in the early
post-inflationary Universe, and these initial conditions completely determine the mass function
of the clusters \cite{cl2}.

In the present Section we consider the BEC dark matter properties as derived
from the observed properties of the galaxy clusters. Galaxy clusters are
complex astrophysical systems, consisting of hundreds to thousands of
galaxies, intergalactic gas, and dark matter, all bound together by gravity.
The basic hypothesis in the study of the galactic clusters is that the
intergalactic hot gas particles are in hydrostatic equilibrium under the
gravity of the dark matter and of the galaxies \cite{RB}. Starting from this
assumption, and by adopting some realistic astrophysical models for the gas
density, the astrophysical properties of the BEC dark matter distribution can be derived
theoretically, thus allowing a full comparison of the model predictions with
observations.

\subsection{Astrophysical parameters of BEC dark matter in galactic clusters}

As a first step in our study we assume that the temperature of the dark
matter in the cluster is low enough so that {\it all bosons} condense to form a
Bose-Einstein Condensate. Since the dark matter particles are bound in
hydrostatic equilibrium by the overall mass distribution of the cluster, it
follows that the dark matter density $\rho _{DM}$ and pressure $P_{DM}$ are
related to the mass of the cluster $M(r)$ by the hydrostatic equilibrium
equation
\begin{equation}  \label{1}
\frac{1}{\rho_{DM}}\frac{dP_{DM}}{dr}=-\frac{GM(r)}{r^2}
\end{equation}
The pressure of the gravitationally bounded BEC dark matter obeys a
polytropic equation of state, given by \cite{BoHa07}
\begin{equation}  \label{2}
P_{DM}=\frac{2\pi\hbar^2a}{m^3}\rho_{DM}^2
\end{equation}
where $m$ and $a$ are the mass and the scattering length of the dark matter
particles in the condensate. Eq.~(\ref{2}) can be derived from the hydrodynamic representation
of the BECs, given by Eqs.~(\ref{euler}) and (\ref{pois}). For the
interstellar gas distribution we adopt King's $\beta$-model, in which the
number density $n_g$ of the hot gas is given by \cite{RB}
\begin{equation}  \label{3}
n_g(r)=n_g(0)\left(1+\frac{r^2}{r_c^2}\right)^{-3\beta/2},
\end{equation}
where $n_g(0)$ is the central number density of the gas, $r_c$ is the
core radius, and $\beta $ is a constant. The gas is assumed to obey the
ideal gas equation of state $P_g=k_Bn_gT_g$, where  $T_g$ is the gas temperature. By assuming again that the hot
gas is in hydrostatic equilibrium in the cluster, similarly to Eq.~(\ref{1}%
), a relation for the pressure of the hot gas as a function of the total
mass of the cluster inside radius $r$ is obtained as
\begin{equation}  \label{4g}
\frac{1}{m_gn_g}\frac{d(n_gk_BT_g)}{d r}=\frac{k_BT_g}{\mu m_p}\left(\frac{1%
}{n_g}\frac{dn_g}{dr}+\frac{1}{T_g}\frac{dT_g}{dr}\right) =-\frac{GM(r)}{r^2}%
,
\end{equation}
where $m_g$ is the mass of the hot gas particle. For the mass of gas
particle we take $m_g={\mu}m_p$ \cite{RB}, where $\mu=0.61$, and $m_p$ is the
proton mass.

When the hot interstellar gas density $\rho _g$ and its temperature profile $%
T_g$ are known from observations, the total mass within a radius $r$ can be
estimated by solving the equation of hydrostatic equilibrium Eq.~(\ref{4g}).
In spherical symmetry we obtain \cite{RB1}
\begin{equation}  \label{mass1}
M(< r) = -\frac{k_BT_gr^2}{G\mu m_p}\left(\frac{1}{\rho _g}\frac{d\rho _g}{dr%
} +\frac{1}{T_g}\frac{dT_g}{dr} \right)=-\frac{k_BT_gr^2}{G\mu m_p}\frac{d}{%
dr}\ln \left[\rho _g(r)T_g(r)\right].
\end{equation}

\subsection{The isothermal gas case}

If we adopt the isothermal condition for the hot gas $T_g=\mathrm{constant}$
from Eq.~(\ref{4g}) we obtain
\begin{equation}  \label{4}
\frac{k_BT_g}{m_gn_g}\frac{d n_g}{dr}=-\frac{GM(r)}{r^2}.
\end{equation}

Combining Eqs.~(\ref{1}), (\ref{3}) and (\ref{4}), we obtain the following
relation between the condensed dark matter density $\rho_{DM}$ and the hot
gas parameters,
\begin{equation}  \label{5}
\frac{4\pi\hbar^2a}{m^3}\frac{d\rho_{DM}}{d r}=\frac{k_BT_g}{m_g}\frac{d}{dr}%
\left(\ln n_g\right).
\end{equation}
By integrating Eq.~(\ref{5}), it follows that the condensate dark matter
density profile in galactic clusters is obtained as
\begin{equation}  \label{6}
\rho_{DM}(r)=\rho_{DM}(0)-\frac{3\beta }{8\pi }\frac{m^3}{m_g}\frac{k_BT_g}{%
\hbar ^2a}\ln\left(1+\frac{r^2}{r_c^2}\right),
\end{equation}
where $\rho_{DM}(0)$ is the central dark matter density.

By introducing the cluster radius via the boundary condition requiring that $%
\rho_{DM}(R)=0$, we obtain the radius of the cluster as a function of the
central dark matter density $\rho_{DM}(0)$ and the gas parameters as
\begin{equation}  \label{7}
R=r_c\sqrt{\exp\left[\frac{8\pi}{3\beta }\frac{m_g}{m^3}\frac{\hbar ^2a}{%
k_BT_g}\rho _{DM}(0)\right]-1},
\end{equation}
or
\begin{equation}
R=r_c\sqrt{\exp\left[10^{-94}\times \frac{1}{\beta}\frac{\left(a/\mathrm{cm}%
\right)}{\left(m/g\right)^3}\frac{1}{\left(T/\mathrm{keV}\right)}\frac{\rho
_{DM}(0)}{10^{-24}\;\mathrm{g/cm^3}}\right]-1}.
\end{equation}
As a function of the cluster radius given by Eq.~(\ref{7}), the condensed
dark matter density distribution in the cluster, Eq.~(\ref{6}), can be
rewritten as
\begin{equation}
\rho_{DM}(r)=\frac{3\beta }{8\pi }\frac{m^3}{m_g}\frac{k_BT_g}{\hbar ^2a}%
\ln\left(\frac{R^2+r_c^2}{r^2+r_c^2}\right), r\leq R.
\end{equation}

Therefore, the total mass profile of the dark matter is
\begin{equation}  \label{8}
M_{DM}(r)=\int_0^r 4\pi\left(r^{\prime }\right)^2\rho_{DM}\left(r^{\prime
}\right)d r^{\prime }=\bar{\rho}r_c^3I\left(\frac{r}{r_c}\right),
\end{equation}
where
\begin{equation}
I\left(\frac{r}{r_c}\right)=\int_0^{r/r_c} {\xi ^2 \ln\frac{1+\xi _0^2}{%
1+\xi ^2}d \xi},
\end{equation}
and
\begin{equation}
\xi _0=\frac{R}{r_c}, \xi =\frac{r}{r_c}, \bar{\rho}=\frac{3\beta }{2}\frac{%
m^3}{m_g}\frac{k_BT_g}{\hbar ^2 a}\label{eq:barrho}
\end{equation}

The function $I\left( r/r_{c}\right) $ can be obtained in an explicit form
as
\begin{equation}
I\left( \frac{r}{r_{c}}\right) =\frac{2}{9}\left( \frac{r}{r_{c}}\right)
^{3}+\frac{1}{3}\left( \frac{r}{r_{c}}\right) ^{3}\ln \left( \frac{1+\text{$%
\xi $}_{0}^{2}}{1+r^{2}/r_{c}^{2}}\right) +\frac{2}{3}\tan ^{-1}\left( \frac{%
r}{r_{c}}\right) -\frac{2}{3}\frac{r}{r_{c}}.
\end{equation}

In both Eqs.~ (\ref{7}) and (\ref{8}), the unknown mass and scattering
length of dark matter particle, $m$ and $a$ appear in the form of the
combination $m^3/a$. In the following we consider a parameter $\lambda$ that
is defined as
\begin{equation}
\lambda=\left(\frac{m}{\text{g}}\right)^3\left(\frac{a}{\text{cm}}%
\right)^{-1},
\end{equation}
and which fully describes the properties of the BEC condensed dark matter.

Remembering Eq.~(\ref{mass}), the observed mass of a cluster is derived as
\cite{RB}
\begin{equation}
M_{obs}(<r)=\frac{3k_B T_g r^3 \beta}{\mu m_p G}\frac{1}{r^2+r_c^2},
\end{equation}
or
\begin{equation}
\frac{M_{obs}(r)}{10^{14}M_{\odot}}=1.086\times 10^{-3}\beta\left(\frac{ T_g%
}{\text{keV}}\right) \frac{\left(r/\mathrm{kpc}\right)^3}{\left(r/\mathrm{kpc%
}\right)^2+\left(r_c/\mathrm{kpc}\right)^2}.
\end{equation}

By assuming that the total mass of the cluster is related to the dark matter
by a relation of the form
\begin{equation}
M_{DM}=k_{DM}\left(M,n_g,T_g,...\right)M_{obs},
\end{equation}
where generally $k_{DM}$ is a function of the total mass of the cluster, of
the gas density, gas temperature etc., we obtain the parameter $\lambda $ of
the condensed dark matter as a function of the global parameters of the
cluster $\left(R,r_c\right)$ as
\begin{equation}
\lambda =2k_{DM}\left(M,n_g,T_g,...\right)\frac{\hbar ^2}{G}\frac{1}{r_c^2}%
\left(\frac{R}{r_c}\right)^3\frac{I^{-1}\left(R/r_c\right)}{%
1+\left(R/r_c\right)^2}.
\end{equation}
or
\begin{equation}\label{eq:isoDefL}
\lambda =3.511\times 10^{-90}\times k_{DM}\left(M,n_g,T_g,...\right)\times
\left(\frac{r_c}{\mathrm{kpc}}\right)^{-2}\times \left(\frac{R}{r_c}\right)^3%
\frac{I^{-1}\left(R/r_c\right)}{1+\left(R/r_c\right)^2}\;\mathrm{g^3/cm}.
\end{equation}

\subsubsection{Numerical Analysis}

The first step in the observational study of the galactic clusters is the
determination of the integrated mass as a function of the radius $r$ \cite{RB}. Once
the integrated mass is known, one must define a physically meaningful
fiducial radius for the mass measurement. There are two such radii  used
by astronomers to interpret observations, denoted as $r_{200}$ and $r_{500}$, respectively. These radii are defined
as the radii where the mean gravitational mass density of the matter $%
\left<\rho _{tot}\right>$ has the values $\left<\rho _{tot}\right> = 200\rho
_{cr}$ and $\left<\rho _{tot}\right> =500\rho _{cr}$, respectively, with $\rho _{cr}$, representing the critical cosmological density
given by $\rho _{cr} (z) = 3H_0^2h^2(z)/8\pi G$, where $h(z)$ is the Hubble
parameter normalized to its local value, i.e., $h^2(z) = \Omega _m (1 + z)^3
+ \Omega _{\Lambda}$,where $\Omega _m$ is the mass density parameter, and $%
\Omega _{\Lambda}$ is the dark energy density parameter, respectively \cite%
{cl1}. $H_0$ is the present day value of the Hubble function, and $z$
denotes the cosmological redshift.

\paragraph{The astrophysical data set}

A generally used pragmatic approach for
determining the virial mass $M_{vir}$ of a galactic cluster is to use $%
r_{200}$ as the outer boundary of the mass distribution in the cluster \cite%
{RB}. The numerical values of the cluster radius $r_{200}$ are in the range $%
r_{200} = 0.85$ Mpc (for the cluster NGC 4636) and $r_{200} = 4.49$ Mpc (for
the cluster A2163). Observations show that a typical value for $r_{200}$ is
approximately 2 Mpc \cite{RB}. The masses corresponding to $r_{200}$ and $%
r_{500}$ are denoted by $M_{200}$ and $M_{500}$, respectively \cite{RB}. In the study
of the gravitational dynamics of galaxy clusters it is usually assumed that $%
M_{vir} \approx M_{200}$ and $R_{vir} \approx r_{200}$, where $R_{vir}$ denotes the
virial radius of the cluster \cite{RB}.

\begin{figure}[tbp]
\centering
\includegraphics[width=0.7\textwidth]{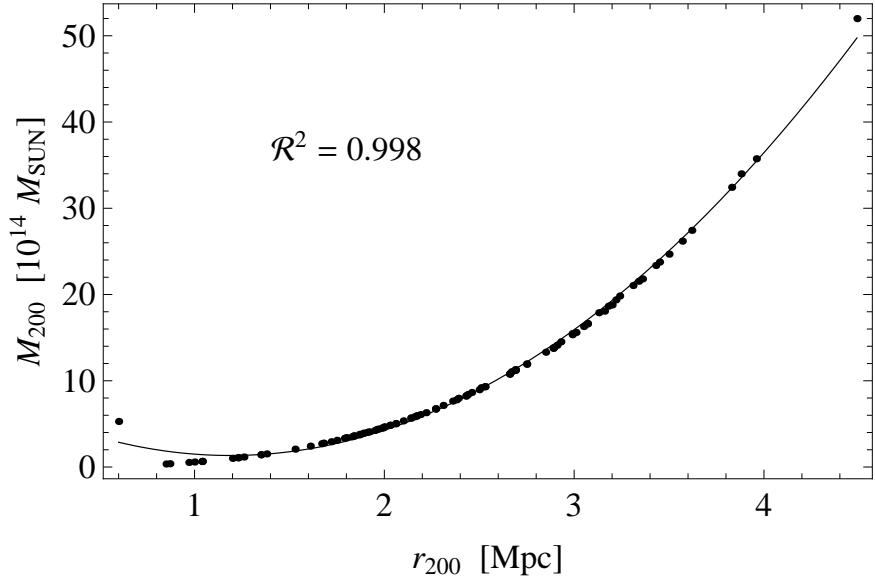}
\caption{Mass-radius relation for 106 clusters \cite{RB}.}\label{fig:M200vsR200}
\end{figure}

\begin{figure}[tbp]
\centering
\includegraphics[width=0.7\textwidth]{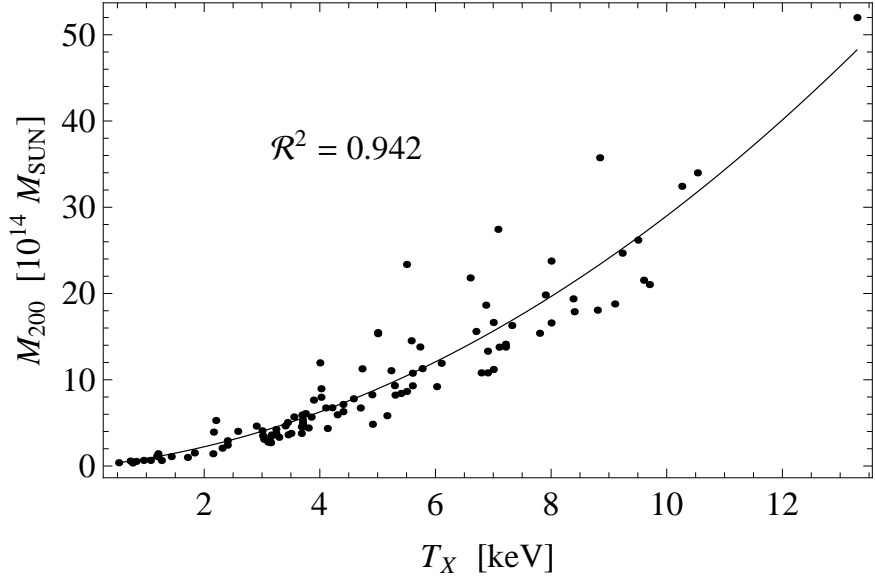}
\caption{Cluster mass-gas temperature relation for 106 clusters \cite{RB}.}\label{fig:M200vsTx}
\end{figure}

In the following we will use the observational data presented in \cite{RB}. The  mass-radius and mass-temperature dependencies of the considered clusters are presented in Figs.~\ref{fig:M200vsR200} and~\ref{fig:M200vsTx}, respectively, where $M_{200}$, $R_{200}$ and $T_X$ are data corresponding to each of the 106 clusters analyzed in~\cite{RB}.

The observational data can be fitted with some simple functions, which allows to obtain the cluster mass-cluster radius relation in the form
\be
\frac{M_{200}}{10^{14}M_{\odot}}=4.439\left(\frac{r_{200}}{{\rm Mpc}}\right)^2-10.545\left(\frac{r_{200}}{{\rm Mpc}}\right)+7.599,
\ee
with a correlation coefficient $\mathcal{R}^2=0.998$. The cluster mass - hot intracluster gas temperature relation can be obtained analytically as
\be
\frac{M_{200}}{10^{14}M_{\odot}}=0.221\left(\frac{T_X}{{\rm keV}}\right)^2+0.690\left(\frac{T_{X}}{{\rm keV}}\right)-0.038,
\ee
with a correlation coefficient $\mathcal{R}^2=0.942$.

In the following we neglect the mass
of the cluster bulge and the cluster disc as their masses ($M_b$, $M_d$ $%
\sim 10^{10}M_{\odot}$) are much smaller than the dark matter mass \cite{Sof}%
. With this in mind, we shall assume that the dark matter accounts for most of the mass (around 90\%) of the cluster.


\paragraph{The physical parameter $\lambda $ of the dark matter particle}

Consider the definition of $\lambda$, given by Eq.~\eqref{eq:isoDefL}, in the context of the observational data provided by analysis of the 106 clusters. For each of these clusters $\lambda$ takes a different value, $\lambda _i$. Following a simple averaging, with the use of the observational data of \cite{RB} we obtain
\be
\langle \lambda \rangle =  \sum _{i=1}^{n} \frac{\lambda _i }{n} = 4.27969\times 10^{-96} \,{\rm g^3/cm},
\ee
for $k_{DM}  = 0.9$, while
\be
\frac{\langle \lambda \rangle }{k_{DM}} = 4.75522 \times 10^{-96} \; {\rm g^3/cm}.
\ee


In a different approach, one can fit the results of Eq.~\eqref{8}, representing the theoretical prediction for the dark matter mass,  for each of the clusters with the observed cluster mass data in~\cite{RB}. This is a straightforward problem once one notices that, with the help of Eq.~\eqref{eq:isoDefL} and the definition of $\bar{\rho}$ from Eq.~\eqref{eq:barrho}, the theoretical expression for the dark matter mass may be written as

\begin{equation*}
M_{DM} = \lambda x, \quad x = \frac{3\beta }{2m_g}\frac{k_BT_g}{\hbar ^2} r_c^3 I\left ( \frac{r}{r_c} \right ).
\end{equation*}

By considering that the observational mass $M_{200}$ is a vector $y$, the problem becomes to find the parameter $\lambda$ which produces the best fit of the equation $\lambda x$ to $y$. The standard approach to solving such a problem is the least squares method~\cite{nash}, i.e. find $\lambda$ which minimises $\sum _i ^n \left ( y_i - \lambda x_i \right ) ^2$. To this end, we define a function
\begin{equation}
Z^2 _M=\sum_{i=1}^n [M_{200}^i-M_{DM}^i(r_{200})]^2
\end{equation}
where $M_{DM}^i(R)$ is obtained from Eq.~(\ref{8}). Please be aware that the standard notation for this function is $R^2$. We chose to denote it $Z^2$ so as to avoid confusion with the different radii used in the paper. The value of $\lambda$ which minimises  the function $Z ^2 _M$ is found to be
\be
\lambda _0 = 1.76624 \times 10^{-96}\;{\rm  g^3/cm}.
\ee

Similarly, two different approaches may be used to find the central density of the dark matter. First, by using Eq.~(\ref{7}) with $R \equiv r_{200}$, we can find, for each cluster (with fixed $\lambda _0$), a value $\rho _{DM}^i(0)$, and thus a mean value
\be
\langle \rho _{DM}(0) \rangle = \sum _{i=1}^{n} {\frac{\rho _{DM}^i(0)}{ n}}=4.86714 \times 10^{-13}\;{\rm  g/cm^3}.
\ee

In order to find the value of $\rho _{DM}(0)$ that best fits the entire data set, we use Eq.~\eqref{7}, and define the functions
\begin{equation}
F _ i = \log \left \{\left (\frac{r_{200}^i}{r_c^i} \right ) ^2 + 1 \right \},\; G_i = \frac{8\pi}{3\beta _i }\frac{m_g}{m^3}\frac{\hbar ^2a}{k_BT_g ^i}\rho _{DM}(0),
\end{equation}
and
\begin{equation}
Z^2 _R = \sum_{i=1}^n \left ( F_i - G_i \right )^2,
\end{equation}
respectively. The value of $\rho _{DM}(0)$ which extremises  $Z^2 _R$ is
\be
\rho _{DM}^{(0)}(0)=1.90934 \times 10^{-13} \;{\rm g/cm^3},
\ee
for fixed $m^3 / a =\lambda _0$.

\begin{figure}[tbp]
\centering
\includegraphics[width=0.7\textwidth]{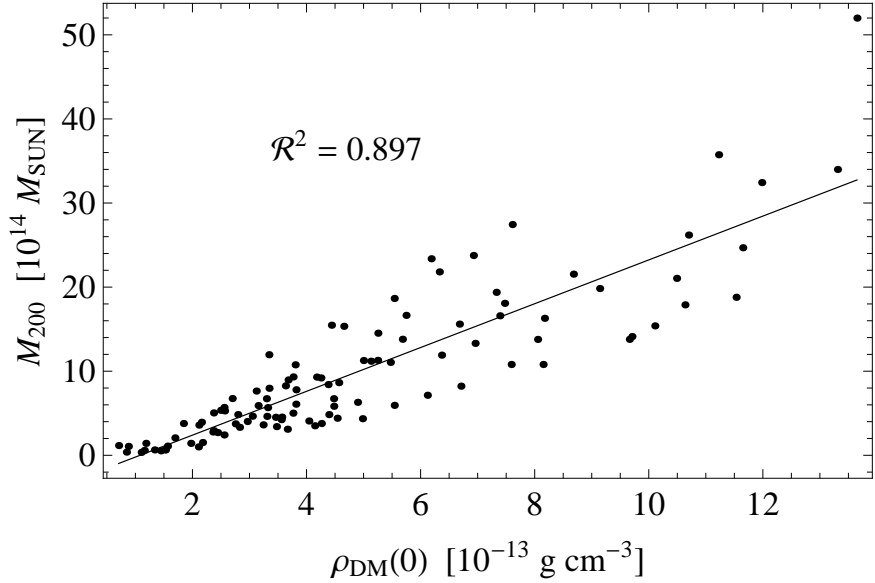}
\caption{Predicted mass-central density relation for 106 clusters, in the case of the isothermal intracluster gas model.}\label{fig:M200vsRho0}
\end{figure}

\begin{figure}[tbp]
\centering
\includegraphics[width=0.7\textwidth]{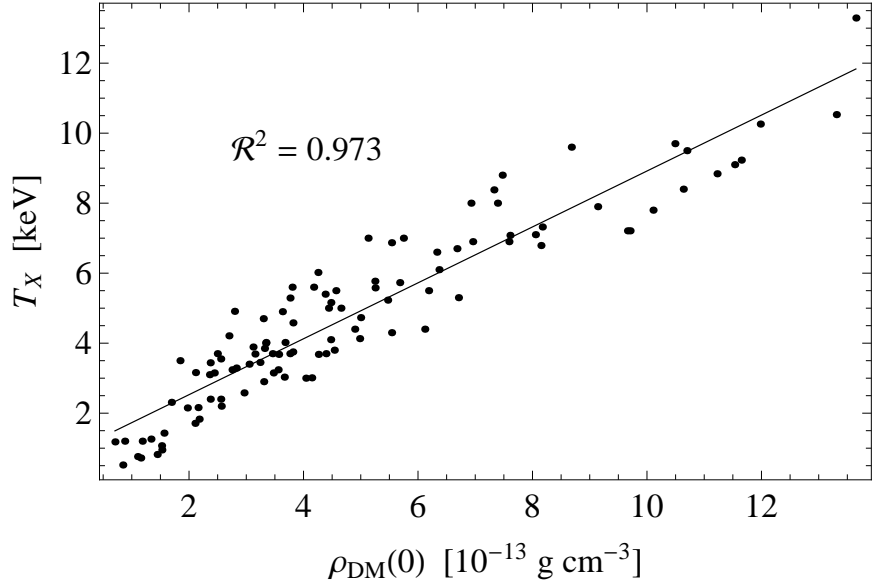}
\caption{Predicted gas temperature-central density relation for 106 clusters, in the case of the isothermal intracluster gas model.}\label{fig:TvsRho0}
\end{figure}

With these results, the plots of the cluster mass-cluster central density and hot gas temperature-cluster central density relations are shown in Figs.~\ref{fig:M200vsRho0} and \ref{fig:TvsRho0}. Fitting an equation of the type $ax + b$ gives the remarkably simple relations
\be
\frac{M_{200}}{10^{14}M_{\odot}}=2.604 \frac{\rho _{DM}(0)}{10^{-13}\;{\rm g/cm^3}} -2.812,
\ee
with a correlation coefficient $\mathcal{R}^2 = 0.897$ for the cluster mass-cluster central density relation, and
\be
\frac{T_X}{{\rm keV}}=0.798\frac{\rho _{DM}(0)}{10^{-13}\;{\rm g/cm^3}} + 0.930,
\ee
with $\mathcal{R}^2 = 0.973$ for the intracluster gas temperature-cluster central density relation, respectively.





\subsubsection{Astrophysical implications: limits on the mass and scattering length of the BEC dark matter particles}





An important characteristic of the dark matter particles is their self-interaction cross-section, defined as
\begin{equation}\label{cross}
\sigma_m=\frac{4\pi a^2}{m}.
\end{equation}

Recent observations \cite{d5}, as well as numerical simulations of the merging galaxy cluster 1E 0657-56 (the Bullet
cluster) \cite{Bul}, have suggested an upper limit ($68\%$ confidence for simulations) for $\sigma_m$ of the order of $\sigma_m<1.25%
\;\text{cm$^2$$\slash$g}=2.23\times 10^{-7}\;\text{fm$^2\slash$eV}$. By combining Eq.~\eqref{eq:isoDefL}, giving the definition of the parameter $\lambda $,  with Eq.~(\ref{cross}), we can obtain the mass and scattering length  of the BEC dark matter particles as
\begin{equation}\label{mass1c}
m=\sqrt[5]{\frac{\lambda _0^2\sigma_m}{4\pi}}<4.4583\times 10^{-39}\;g=1.7638\times 10^{-3}\; {\rm meV},
\end{equation}
\begin{equation}\label{a1n}
a=\sqrt[5]{\frac{\lambda _0\sigma_m^3}{(4\pi)^3}}<1.7702\times 10^{-20}\;{\rm cm}=1.7702\times 10^{-7}\;{\rm fm},
\end{equation}
where for the numerical estimations we have considered the upper limits of $\sigma_m$ and $\lambda$. Thus, the presented numerical values give the upper limits for the mass and scattering length of the dark matter particles.
The upper limit of BEC dark matter particle mass is a little bit smaller than the region that
QCD predicts for the mass of the axion, $6.0\;\mu\text{eV}<m<2.0\;\text{meV}$ \cite{axion}. This
upper limit is also smaller than the result in \cite{Beck} that the
axion mass is of the order of $m=0.11\;\text{meV}$. The value of the scattering length
obtained from the cross section of the self-interacting dark matter and zero-temperature approximation, is much
smaller than the value of $a=10^4-10^6\;\text{fm}$ obtained in laboratory
experiments on atomic BECs \cite{exp}.

\section{The effect of the intracluster hot gas temperature variation on
the BEC dark matter density profile}\label{sectn}

In the previous Section we have considered that the intracluster gas is
isothermal, $T_g=\mathrm{constant}$, and there are no temperature variations
in the gas temperature. However, a large number of astrophysical investigations have
shown that the gas temperature structure on Mpc scales is highly complex and
non-isothermal \cite{Mark,Lok,RB1,Vikh,Evr,Dur,Gas,Sof}. Through a detailed analysis of the
observational data of a sample of hot nearby clusters it was found that the
gravitating mass within 1 and 6 times core radius are approximately within
1.35 and 0.7 times that the isothermal $\beta$-model estimates \cite{Mark}.
From the analysis of the temperature of 13 galaxy clusters a scaled
temperature profile of the form $T\slash\langle T \rangle= 1.07, 0.035<r%
\slash r_{180}<0.125$, $T\slash\langle T \rangle= 1.22-1.2r\slash r_{180},
0.125<r\slash r_{180}<0.6$ was obtained in \cite{Vikh}, where $r_{180}=2.74%
\text{Mpc}(\langle T \rangle \slash 10\text{keV})^{1 \slash 2}$ is the
cluster's virial radius \cite{Evr}. As for the central region ($r\slash %
r_{180}<0.035$) of the cluster, the temperature profiles scatter since
non-gravitational processes such as radiative cooling and energy output are
important. Hence, in order to obtain a realistic description of the galaxy
cluster properties one should take into account in the left hand side of
Eq.~(\ref{4g}) the presence of the temperature gradients in the hot
intracluster gas.

On the other hand in \cite{Chu} it was suggested that the temperature
variations may only amount to 20-30$\%$ ( $T_g\,\mathrm{d} n_g\slash \,%
\mathrm{d} r \sim 4-5$ times greater than $n_g\,\mathrm{d} T_g\slash \,%
\mathrm{d} r$), thus having little effect on the astrophysical parameters of
the cluster.

In order to examine the effect of the intracluster gas temperature
variations on the Bose-Einstein Condensate dark matter properties we adopt a
very simple model for the temperature variations, described by the equation
\cite{Mark,Lok},
\begin{equation}
T_{g}(r)=\frac{T_{0}}{\left[ 1+\left( r/r_{c}\right) ^{2}\right] ^{\delta }},
\label{Tg}
\end{equation}%
where $T_{0}=T(0)$ is the gas temperature at the cluster center, and $\delta
=3\beta (\gamma -1)/2$, where $\beta =2/3$ and $\gamma =1.24_{-0.12}^{+0.20}$%
. Therefore in the presence of gas temperature gradients the BEC dark matter
profile is given by the equation
\begin{equation}
\frac{4\pi \hbar ^{2}a}{m^{3}}\frac{d\rho _{DM}(r)}{dr}=\frac{k_{B}T_{g}(r)}{%
\mu m_{p}}\frac{d}{dr}\left[ \ln n_{g}(r)T_{g}(r)\right] .
\end{equation}%
By taking into account the explicit expressions of the gas particle density
and of the temperature, given by Eqs.~(\ref{3}) and (\ref{Tg}),
respectively, we obtain first
\begin{equation}
\frac{4\pi \hbar ^{2}a}{m^{3}}\frac{d\rho _{DM}(r)}{dr}=-\frac{3\beta \gamma
k_{B}T_{0}}{\mu m_{p}r_{c}^{2}}r\left( 1+\frac{r^{2}}{r_{c}^{2}}\right)
^{-3\beta (\gamma -1)/2-1},
\end{equation}%
giving the BEC dark matter density profile as
\begin{equation}
\rho _{DM}(r)=C+\frac{\gamma }{4\pi (\gamma -1)\mu }\frac{m^{3}}{m_{p}}\frac{%
k_{B}T_{0}}{\hbar ^{2}a}\left( 1+\frac{r^{2}}{r_{c}^{2}}\right) ^{-3\beta
(\gamma -1)/2},
\end{equation}%
where $C$ is an arbitrary integration constant. Since
\begin{equation}
\rho _{DM}(0)=C+\frac{\gamma m^{3}k_{B}T_{0}}{4\pi (\gamma -1)\mu m_{p}\hbar
^{2}a},
\end{equation}
it follows that finally the dark matter density profile is obtained as
\begin{equation}
\rho _{DM}(r)=\rho _{DM}(0)-\frac{\gamma }{4\pi (\gamma -1)\mu }\frac{m^{3}}{%
m_{p}}\frac{k_{B}T_{0}}{\hbar ^{2}a}\left[ 1-\left( 1+\frac{r^{2}}{r_{c}^{2}}%
\right) ^{-3\beta (\gamma -1)/2}\right] .  \label{densT}
\end{equation}

The radius of the cluster is obtained from the condition $\rho _{DM}(r)=0$,
giving
\begin{equation}\label{eq:nonisoRad}
R=r_{c}\sqrt{\left[ 1-\frac{4\pi (\gamma -1)\mu }{\gamma }\frac{m_{p}}{m^{3}}%
\frac{\hbar ^{2}a}{k_{B}T_{0}}\rho _{DM}(0)\right] ^{-2/\left [3\beta (\gamma -1)\right ]}-1%
},
\end{equation}%
and
\begin{equation}
\rho _{DM}(r)=\frac{\gamma }{4\pi (\gamma -1)\mu }\frac{m^{3}}{m_{p}}\frac{%
k_{B}T_{0}}{\hbar ^{2}a}\left[ \left( 1+\frac{r^{2}}{r_{c}^{2}}\right)
^{-3\beta (\gamma -1)/2}-\left( 1+\frac{R^{2}}{r_{c}^{2}}\right) ^{-3\beta
(\gamma -1)/2}\right] ,
\end{equation}
respectively.

The total dark matter mass of the cluster is obtained from $M_{DM}(r)=4\pi
\int_{0}^{r}\rho _{DM}(r^{\prime })r^{\prime 2}dr^{\prime }$, and is
represented by the relation
\begin{equation}\label{eq:nonisoMdm}
M_{DM}(r)=\frac{\gamma }{3(\gamma -1)\mu }\frac{m^{3}}{m_{p}}\frac{k_{B}T_{0}%
}{\hbar ^{2}a}r^{3}\left[ \,_{2}F_{1}\left( \frac{3}{2},\frac{3}{2}\beta
(\gamma -1);\frac{5}{2};-\frac{r^{2}}{r_{c}^{2}}\right) -\left( 1+\frac{R^{2}%
}{r_{c}^{2}}\right) ^{-\frac{3}{2}\beta (\gamma -1)}\right] ,
\end{equation}%
where the hypergeometric function $\,_{2}F_{1}(a,b;c;z)$ is defined as $%
\,_{2}F_{1}(a,b;c;z)=\sum_{k=0}^{\infty }\left( a_{k}\right) \left(
b_{k}\right) \left( c_{k}\right) z^{k}/k!$.

With the use of Eq.~(\ref{mass}) we obtain the total mass of the cluster in
the present varying gas temperature model as
\begin{equation}
M(r)=\frac{3\beta \gamma }{\mu m_{p}}\frac{k_{B}T_{0}}{G}\frac{1}{r_{c}^{2}}%
r^{3}\left( 1+\frac{r^{2}}{r_{c}^{2}}\right) ^{-3\beta (\gamma -1)/2-1}.
\end{equation}

By assuming that the BEC dark matter mass is proportional to the total mass,
with the proportionality coefficient denoted again by $k_{DM}=k_{DM}\left(M,T_0,..\right)$, we obtain for the
parameter $\lambda $ describing the condensate physical properties the expression
\begin{equation}\label{eq:nonisoDefL}
\lambda =4.5 k_{DM}\beta \left( \gamma -1\right) \frac{\hslash ^{2}}{G}\frac{1}{%
r_{c}^{2}}
\frac{\left( 1+R^{2}/r_{c}^{2}\right)
^{-3\beta (\gamma -1)/2-1}}{ _{2}F_{1}\left( 3/2,3\beta (\gamma
-1)/2;5/2;-R^{2}/r_{c}^{2}\right)-\left( 1+\frac{R^{2}}{r_{c}^{2}}\right) ^{-3\beta (\gamma -1)/2}}.
\end{equation}

Applying the same type of numerical procedure as presented for the isothermal case, with $\gamma = 1.24$, Eq.~\eqref{eq:nonisoDefL} provides
\be
\langle \lambda \rangle = 3.65389 \times 10^{-96} \;{\rm g^3/cm},
 \ee
for $k_{DM} = 0.9$ and
\be
\frac{\langle \lambda \rangle }{k_{DM}} = 4.05987 \times 10^{-96} \;{\rm g^3/cm},
\ee
respectively. A fit of the observational data with respect to Eq.~\eqref{eq:nonisoMdm} leads to a value of
\be
\lambda _1 = 4.20896 \times 10^{-96}\;{\rm  g^3/cm},
\ee
 which extremises the $Z ^2 _M$ function.

By replacing $R$ with $r_{200}$ from Eq.~\eqref{eq:nonisoRad} one can extract the value of $\rho _{DM}(0)$ for each cluster. Thus we  obtain $\langle \rho _{DM}(0)\rangle = 8.19759 \times 10^{-13} \;{\rm g/cm^3}$. To obtain a value of the central density that is a best fit for the entire set of data, we use Eq.~\eqref{eq:nonisoRad}, and define the functions
\begin{equation}
F _ i = \left [\left (\frac{R_{200}^i}{r_c^i} \right ) ^2 + 1 \right ] ^{-3\beta _i (\gamma -1)/2},\; G_i = 1-\frac{4\pi (\gamma -1)\mu }{\gamma }\frac{m_{p}}{m^{3}}%
\frac{\hbar ^{2}a}{k_{B}T_i}\rho _{DM}(0),
\ee
and
\begin{equation}
Z^2 _R = \sum_{i=1}^n \left ( F_i - G_i \right )^2,
\end{equation}
respectively.
The value of $\rho _{DM}(0)$ which extremises $Z^2 _R$ is $\rho _{DM}^{(1)}(0)=4.10184 \times 10^{-13} \;{\rm g/cm^3}$ for fixed $m^3 / a=\lambda _1$.

With these results, the plots of the cluster mass-central density and hot gas temperature-central density relations are shown in Figs.~\ref{fig:M200vsRho0-noniso} and \ref{fig:TvsRho0-noniso}. Fitting an equation of the type $ax + b$, produces
\be
\frac{M}{10^{14}M_{\odot}}=1.728 \frac{\rho _{DM}(0)}{10^{-13}\;{\rm g/cm^3}}-4.306 ,
\ee
 with $\mathcal{R}^2= 0.921$ for the cluster mass-cluster central density relation, and
 \be
\frac{T_X}{{\rm keV}}= 5.358 \frac{\rho _{DM}(0)}{10^{-13}\;{\rm g/cm^3}} + 0.425,
\ee
 with $\mathcal{R}^2 = 0.991$ for the hot gas temperature-cluster central density relation.

\begin{figure}[tbp]
\centering
\includegraphics[width=0.7\textwidth]{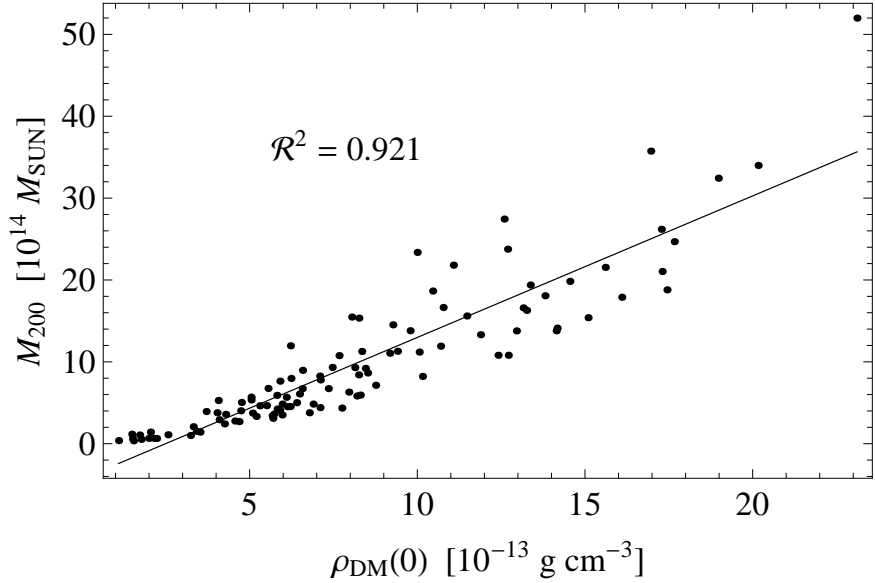}
\caption{Cluster mass-central density relation for 106 clusters, for the non-isothermal intracluster gas model.}\label{fig:M200vsRho0-noniso}
\end{figure}

\begin{figure}[tbp]
\centering
\includegraphics[width=0.7\textwidth]{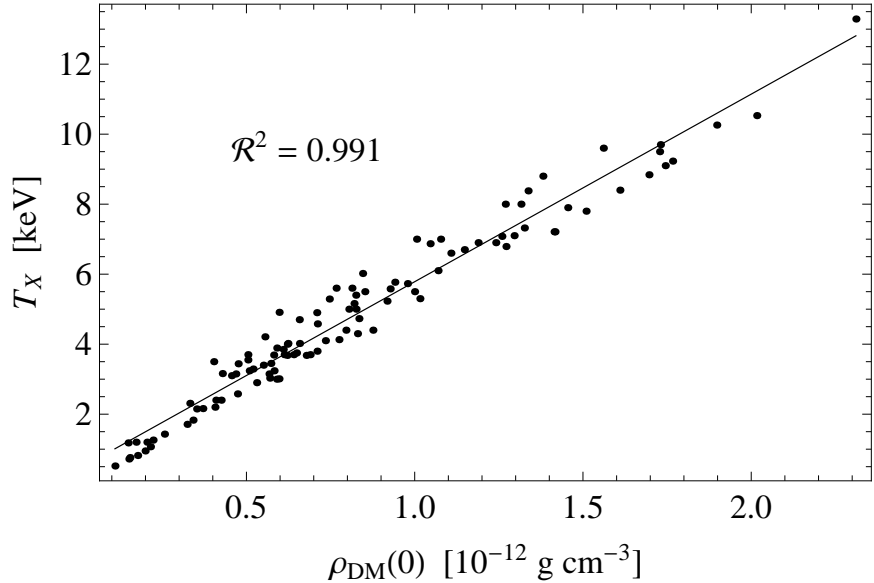}
\caption{Gas temperature-central density relation for 106 clusters, for the non-isothermal intracluster gas model.}\label{fig:TvsRho0-noniso}
\end{figure}




A simple qualitative estimate of the effect of the gas temperature on the
dark matter profile can be obtained by considering an upper limit in the
relation between the temperature gradient and the density gradient in the
gas. By assuming that the contribution of the temperature gradient is
approximately as much as half of the one of the number density gradient, we
obtain
\begin{equation}
n_g\frac{dT_g}{dr}\sim50\%T_g\frac{d n_g}{dr},
\end{equation}
and the dark matter density becomes
\begin{equation}
\rho_{DM}(r)=1.5\times\frac{3k_B T_g\beta m^3}{8\pi\hbar^2m_ga}\ln{\left(%
\frac{R^2+r_c^2}{r^2+r_c^2}\right)}
\end{equation}

Still, detailed temperature profiles and other non-gravitational factors in
the core regions are needed to derive a more accurate density profile of
dark matter.

To sum up, due to a lack of knowledge about the temperature profiles in most
clusters and on the role of the important effects of non-gravitational
factors on the cluster hot gas, there are still many uncertainties in the
derivation of the dark matter density profile by using gravitational methods.

\section{The Finite Temperature Bose-Einstein Condensate dark matter model}\label{sect3}

Unlike the zero temperature BEC dark matter model, in which all bosons stay in the condensed
phase, in the finite temperature BEC model, at least, a fraction of the
particles can be found in the thermal cloud (non-condensed) phase). According
to a fundamental relation of statistical physics, the ratio of the number of the condensed
particles $N_0$ and the total particle number $N$ is given by the relation \cite{Za99, Gr}
\begin{equation}
\frac{\left \langle N_0 \right \rangle}{N}=1-\left(\frac{T}{T_{tr}}%
\right)^\gamma
\end{equation}
where $\gamma=3\slash 2$ in ideal gas BEC and $\gamma=3$ in gas BEC trapped
by harmonic potential. When $T=0$ all the particles are in the condensed
phase. In the following we will consider the effects of the finite
temperature of the BEC dark matter on the astrophysical properties of the
galactic clusters.

\subsection{Astrophysical parameters of the finite temperature BEC dark
matter in the presence of a thermal cloud}

In the following we denote by $\left( \rho _{c},p_{c}\right) $ and $\left(
\tilde{\rho},\tilde{p}\right) $ the energy densities and pressures of the
dark matter in the condensed and non-condensed (thermal cloud) phases.
Moreover, we introduce the dimensionless condensate density $\theta (r)$,
related to the condensate density by the relation
\begin{equation}\label{theta}
\rho _{c}(r)=\rho _{tr}^{2\slash3}\frac{m^{1\slash3}}{a}\theta (r)=\rho
_{tr}\kappa \theta (r),
\end{equation}%
where we have denoted $\kappa =m^{1/3}/\rho _{tr}^{1/3}a$. In the following we also denote $t=T/T_{tr}$.

By considering that both the condensed and the non-condensed bosonic phases
are in chemical potential equilibrium, the total energy density $\rho
_{DM}=\rho _{c}+\tilde{\rho}$ and the total pressure $p_{DM}=p_{c}+%
\tilde{p}$ can be expressed in the form of power expansions in the
dimensionless condensate density as \cite{Har2}
\begin{equation*}
\rho _{DM}\left( t,\theta \right) =\rho _{tr}\Bigg[\kappa \theta
+t^{3/2}-2.642t\sqrt{\theta }+2.120t^{1/2}\theta -0.572t^{-1/2}\theta
^{2}+0.088t^{-3/2}\theta ^{3}\Bigg],
\end{equation*}%
\begin{eqnarray}  \label{press}
p_{DM}\left( t,\theta \right) &=&p_{tr}\Bigg[0.513t^{5/2}-3.793t^{3/2}\theta
+0.004t\theta ^{3/2}+\left( 1.896\kappa +4.021\sqrt{t}\right) \theta ^{2}-
\notag \\
&&\frac{2.139\theta ^{3}}{\sqrt{t}}+\frac{0.504\theta ^{4}}{t^{3/2}}, \Bigg ]%
,
\end{eqnarray}%
where we have denoted
\begin{equation}
p_{tr}=\rho _{tr}\frac{k_{B}T_{tr}}{m}.
\end{equation}

Then, by assuming that both non-condensate and condensate dark matter are
bound in hydrostatic equilibrium by the total mass profile, we obtain
\begin{equation}
\frac{1}{\rho _{DM}}\frac{dp_{DM}}{dr}=\frac{k_{B}T_{g}}{m_{g}}\frac{d\left(
\ln n_{g}\right) }{dr}=-\frac{3\beta k_{B}T_{g}}{m_{g}r_{c}^{2}}\frac{r}{%
\left( 1+r^{2}/r_{c}^{2}\right) },
\end{equation}%
where we have assumed that the intracluster gas is isothermal, $T_{g}=%
\mathrm{constant}$. In the following we introduce the dimensionless variable
$\xi =r/r_{c}$, and we denote
\begin{equation}
\alpha =\frac{3\beta }{\mu }\frac{m}{m_{p}}\frac{T_{g}}{T_{tr}}.
\end{equation}

Therefore the equation describing the finite temperature bosonic dark matter
profile in a galactic cluster is given by
\begin{eqnarray}\label{eqf1}
\hspace{-0.5cm}\frac{2.016\theta ^{3}+\left( 3.792\kappa t^{3/2}+8.04t^{2}\right) \theta
+0.006t^{5/2}\sqrt{\theta }-3.793t^{3}-6.417t\theta ^{2}}{0.088\theta
^{3}+kt^{3/2}\theta -2.642t^{5/2}\sqrt{\theta }+t^{3}+2.12t^{2}\theta -0.572%
\sqrt{t}\theta ^{2}}\frac{d\theta }{d\xi }=
-\frac{\alpha \xi }{1+\xi ^{2}}.
\end{eqnarray}%

Eq.~(\ref{eqf1}) must be integrated with the initial condition $\theta
(0)=\theta _{0}$, a condition which gives the value of the Bose-Einstein
condensed dark matter at the galactic cluster's center. The mass
distribution of the dark matter can be obtained from the equation
\begin{equation}
\frac{dm}{d\xi }=\xi ^{2}\Bigg[\kappa \theta +t^{3/2}-2.642t\sqrt{\theta }%
+2.120t^{1/2}\theta -0.572t^{-1/2}\theta ^{2}+0.088t^{-3/2}\theta ^{3}\Bigg],
\label{eqf2}
\end{equation}%
where
\begin{equation}
m(\xi )=\frac{M(\xi )}{4\pi r_{c}^{3}\rho _{tr}}.
\end{equation}%
The initial condition for Eq.~{\ref{eqf2}) is $m(0)=0$. The variations of
the total density of the finite temperature dark matter, of its pressure,
and of the dark matter mass distribution in the cluster are presented, for
fixed $\kappa $, $\alpha $, and $\theta _{0}$, and for different values of $t
$, in Figs.~\ref{fig1}-\ref{fig3}. }

\begin{figure}[tbp]
\centering
\includegraphics[width=0.7\textwidth]{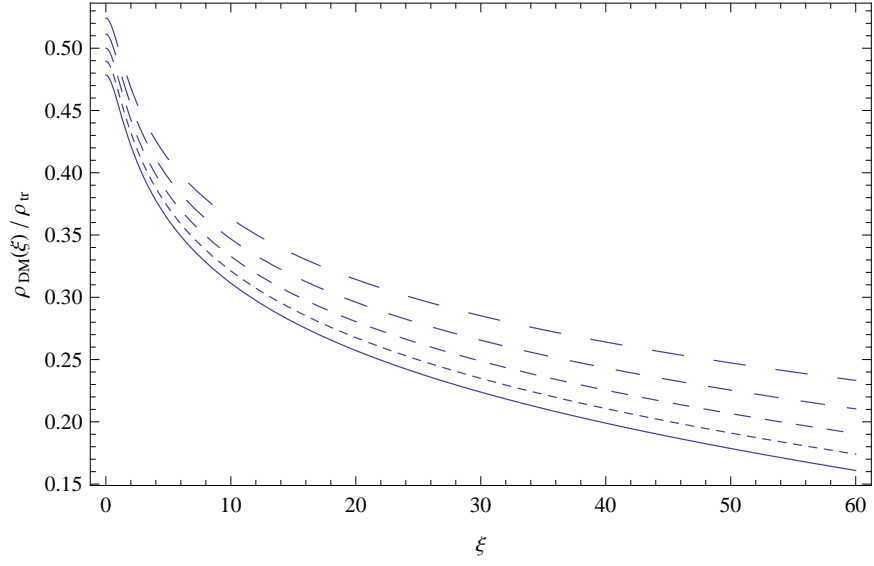}
\caption{Variation of the finite temperature bosonic dark matter
dimensionless density profile $\protect\rho _{DM}/\protect\rho _{tr}$ for $%
\protect\kappa =5$, $\protect\alpha =0.06$, and for different values of $%
t=T/T_{tr}$: $t=0.15$ (solid curve), $t=0.20$ (dotted curve), $t=0.25$
(short dashed curve), $t=0.30$ (dashed curve), and $t=0.35$ (long dashed
curve), respectively. The central value of the Bose-Einstein Condensed dark
matter is $\protect\theta (0)=0.10$. }
\label{fig1}
\end{figure}

\begin{figure}[tbp]
\centering
\includegraphics[width=0.7\textwidth]{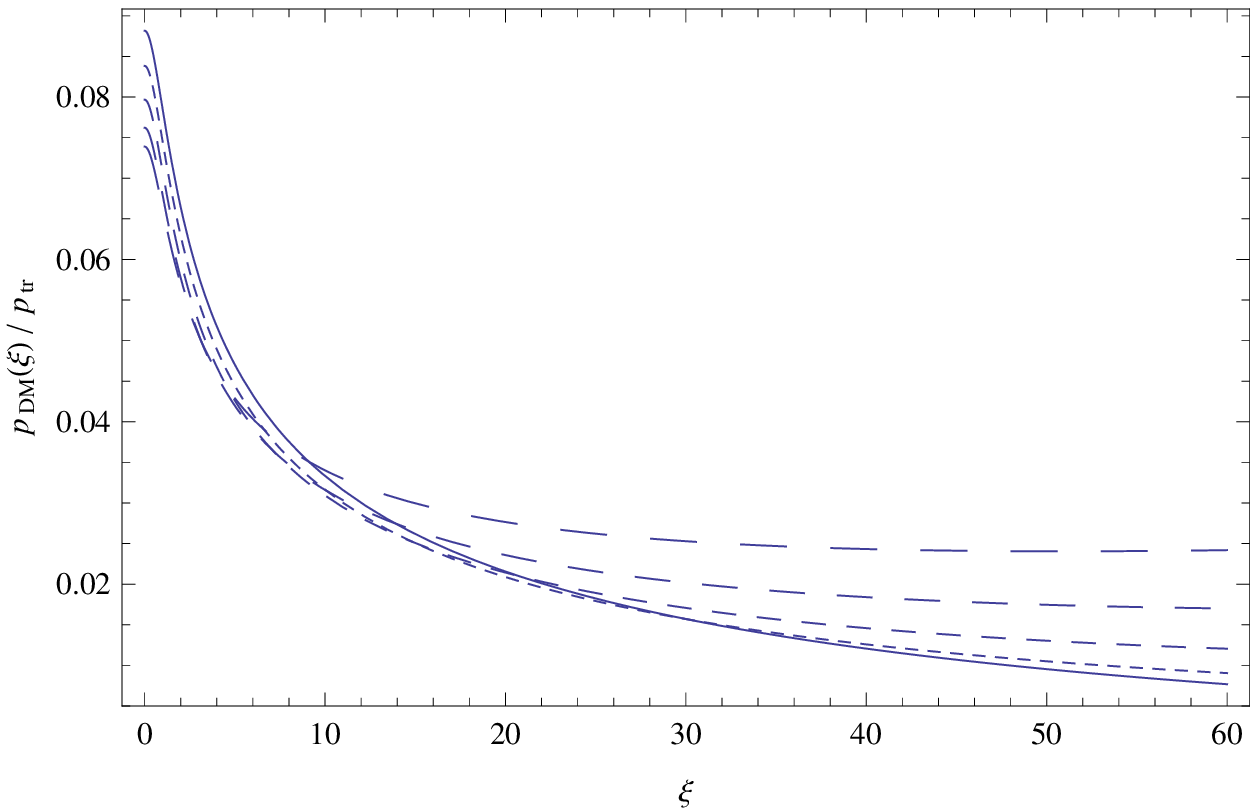}
\caption{Dimensionless pressure profile $p _{DM}/p _{tr}$ of the finite
temperature bosonic dark matter for $\protect\kappa =5$, $\protect\alpha %
=0.06$, and for different values of $t=T/T_{tr}$: $t=0.15$ (solid curve), $%
t=0.20$ (dotted curve), $t=0.25$ (short dashed curve), $t=0.30$ (dashed
curve), and $t=0.35$ (long dashed curve), respectively. The central value of
the Bose-Einstein Condensed dark matter is $\protect\theta (0)=0.10$. }
\label{fig2}
\end{figure}

\begin{figure}[tbp]
\centering
\includegraphics[width=0.7\textwidth]{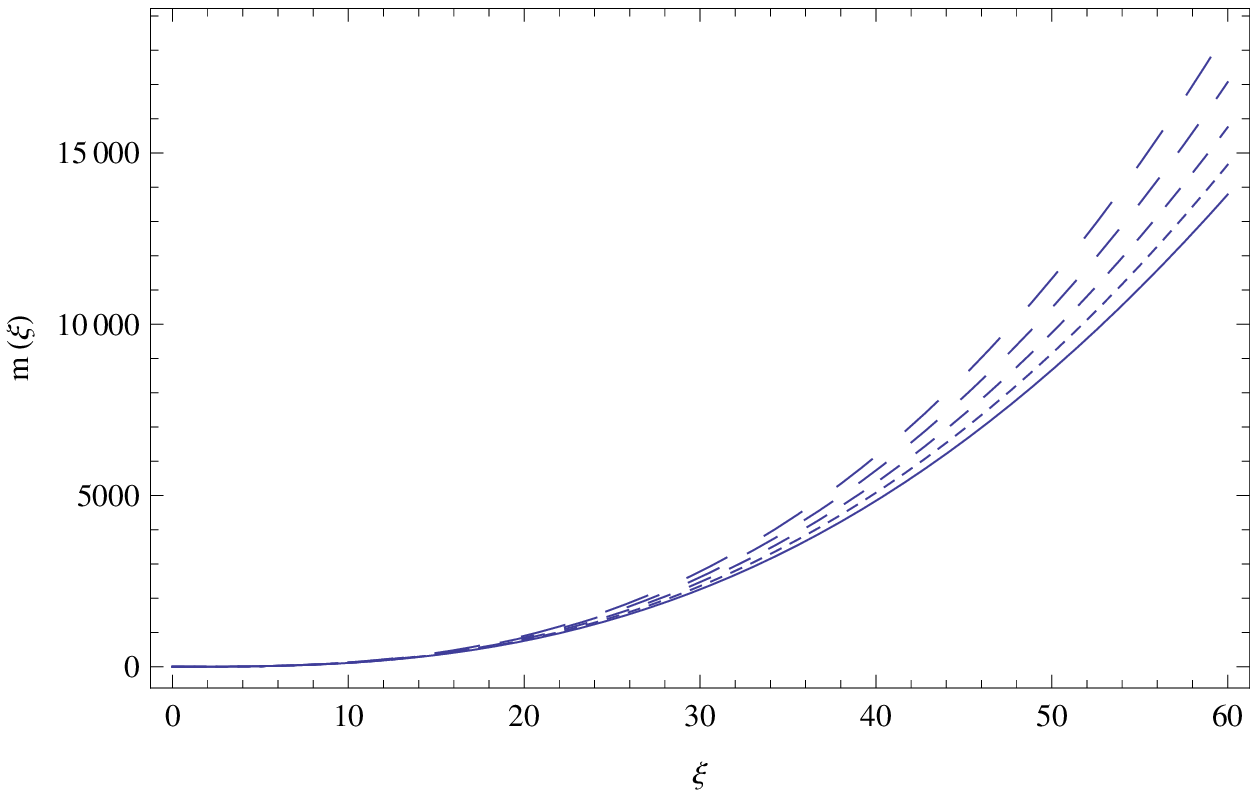}
\caption{Dimensionless mass profile $m(\protect\xi )$ of the finite
temperature bosonic dark matter for $\protect\kappa =5$, $\protect\alpha %
=0.06$, and for different values of $t=T/T_{tr}$: $t=0.15$ (solid curve), $%
t=0.20$ (dotted curve), $t=0.25$ (short dashed curve), $t=0.30$ (dashed
curve), and $t=0.35$ (long dashed curve), respectively. The central value of
the Bose-Einstein Condensed dark matter is $\protect\theta (0)=0.10$. }
\label{fig3}
\end{figure}

\subsection{Finite temperature BEC dark matter: neglecting the effect of the thermal excitations}

In the following we consider a simplified finite temperature dark matter BEC model, in which we assume that the effects of the thermal excitations can be neglected. Thus, we assume that $\rho _c>>\tilde{\rho}$, and $p_c>>\tilde{p}$. These conditions imply that the dark matter temperature is close to the absolute zero temperature. Moreover, we assume that the BEC dark matter is isothermal, that is, its temperature $T$ is a constant inside the cluster. In terms of the dimensionless density $\theta $, defined by Eq.~(\ref{theta}), from Eq.~(\ref{pc}) we obtain for the thermodynamic pressure of the finite temperature BEC dark matter the expression
\bea
p_c\left(T,\theta \right)&=&\rho _{tr}\frac{k_{B}T_{tr}}{m}\Bigg[1.896\kappa \theta ^2
-6.680\left(\frac{T}{T_{tr}}\right)\theta ^{3/2}+8.043\left(\frac{T}{T_{tr}}\right)^{1/2}\theta ^{2}-\nonumber\\
&&2.863\left(\frac{T}{T_{tr}}\right)^{-1/2}\theta ^{3}+
 0.504\left(\frac{T}{T_{tr}}\right)^{-3/2}\theta ^{4}\Bigg].
\eea
The ratio $p_c/p_0$ for the condensate, where $p_0$ is the central pressure, is determined by the numerical value of the dimensionless parameter
$\kappa =m^{1/3}/\rho _{tr}^{1/3}a$. By assuming that the intracluster gas is isothermal with $T_g={\rm constant}$, and in gravitational equilibrium inside the cluster, we obtain the basic equation describing the density distribution of the finite temperature BEC dark matter as
\be
\frac{k_BT_g}{\mu m_p}\frac{d}{dr}\ln n_g=\frac{1}{\rho _c}\frac{dp_c}{dr},
\ee
giving immediately after integration
\bea\label{eqfinn}
C-\frac{3\beta k_BT_g}{2\mu m_p}\ln \left(1+\frac{r^2}{r_c^2}\right)&=&\frac{k_BT_{tr}}{\kappa m}\Bigg[-20.04t\theta ^{1/2}+\left(3.792\kappa +16.08\sqrt{t}\right)\theta -4.294t^{-1/2}\theta ^2+\nonumber\\
&&0.672t^{-3/2}\theta ^3\Bigg],
\eea
respectively, where $C$ is an arbitrary constant of integration, which is determined by the central density of the cluster. Since $\theta <<1$, in the limit of finite $t$ the last two terms in Eq.~(\ref{eqfinn}) can be neglected as being negligibly small. In the zeroth order approximation of the zero temperature limit of the BEC dark matter we obtain first
\be
\theta ^{(0)}(r)=\theta ^{(0)}(0)-\frac{3\beta }{7.584\mu}\frac{m}{m_p}\frac{T_g}{T_{tr}}\ln \left(1+\frac{r^2}{r_c^2}\right),
\ee
where $\theta ^{(0)}(0)=mC/3.792k_BT_{tr}$ is the central value of the dimensionless density of the cluster. This relation also gives the value of the integration constant $C$  in terms of the central value of the dimensionless density $\theta $. At this moment we can define the radius $R$ of the cluster from the condition $\theta ^{(0)}(R)=0$, which gives for the central value of $\theta $ the expression $\theta ^{(0)}(0)= \left(3\beta /7.584\mu\right)\left(m/m_p\right)\left(T_g/T_{tr}\right)\ln \left(1+R^2/r_c^2\right)$.

 In the first order of approximation we have
\be
\theta (r)=\theta ^{(0)}(r)+\theta ^{(1)}(r),
\ee
where $\theta ^{(1)}(r)<<\theta ^{(0)}(r)$, $\forall r\in [0,R]$. After substitution in Eq.~(\ref{eqfinn}) we obtain for $\theta ^{(1)}(r)$ the expression
\be
\theta ^{(1)}(r)=\frac{20.04t-16.08\sqrt{t}}{3.792\kappa-10.02t +16.08\sqrt{t}}\theta _0\approx \left[-\frac{4.24051 \sqrt{t}}{\kappa }+\frac{(5.28481 \kappa +17.9819) t}{\kappa ^2}\right]\theta _0.
\ee
Therefore for the finite temperature Bose-Einstein Condensate dark matter density and mass distribution we obtain the expressions
\be
\rho_{DM}^{(T)}(r,T)\approx\frac{3\beta }{8\pi }\frac{\lambda }{m_g}\frac{k_BT_g}{\hbar ^2}%
\left[1-\frac{4.24051 \sqrt{t}}{\kappa }+\frac{(5.28481 \kappa +17.9819) t}{\kappa ^2}\right]
\ln\left(\frac{R^2+r_c^2}{r^2+r_c^2}\right), r\leq R,
\ee
and
\begin{equation}
M_{DM}^{(T)}(r,T)\approx \bar{\rho}_Tr_c^3I\left(\frac{r}{r_c}\right),\label{eq:Mdm}
\end{equation}
respectively, where
\begin{equation}\label{rhoT}
 \bar{\rho}_T=\frac{3\beta }{2}\frac{%
m^3}{m_g}\frac{k_BT_g}{\hbar ^2 a}\left[1-\frac{4.24051 \sqrt{t}}{\kappa }+\frac{(5.28481 \kappa +17.9819) t}{\kappa ^2}\right].
\end{equation}

In the present order of approximation the radius $R$ of the cluster is not affected by the physical effects related to the finite temperature of the BEC dark matter.

By fixing the value of $\lambda $ to the one obtained as a best fit for the case of zero temperature BEC in an isothermal gas, $\lambda _0 = 1.76624 \times 10^{-96}\;{\rm g^3/cm}$ we explore the parameter space $\{t,\kappa\}$ needed to extremise the values of the function $Z _M ^2$, defined in this case for the dark matter mass as given by Eq.~(\ref{eq:Mdm}). A plot of the variation of these parameters for two values of $\lambda $, $\lambda =\lambda _0$ and $\lambda =\lambda _0/2$, respectively, is shown in Fig.~\ref{fig:diffl}; all points in the plot produce the same minimal value for $Z _M^2$.

\begin{figure}[tbp]
\centering
\includegraphics[width=0.7\textwidth]{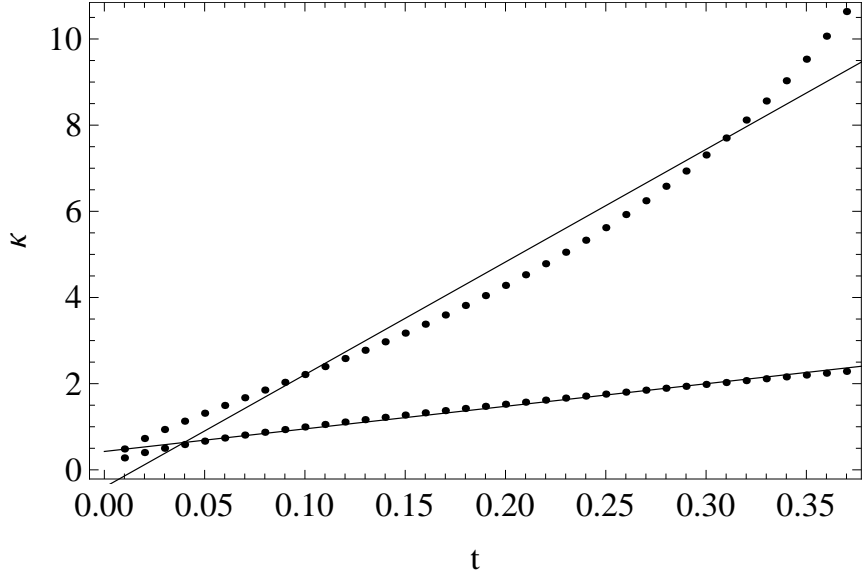}
\caption{Parameter space for the pair $\{t, \kappa\}$, with fixed $\lambda _0$ (upper line) and $\lambda _0 /2$ (lower line).}
\label{fig:diffl}
\end{figure}


In the case of Fig.~\ref{fig:diffl}, a fit with a linear equation produces
\be
\kappa =26.147 t -0.405,
\ee
with $\mathcal{R}^2= 0.991$ for the upper line, and
\be
\kappa = 5.239 t + 0.426,
\ee
 with $\mathcal{R}^2 = 0.998$ for the lower line.




\section{Discussion and Final remarks}\label{sect4}

In the present paper we have provided some preliminary tests at the galactic cluster scale of the hypothesis that dark matter does exist in the form of
a Bose-Einstein Condensate. By assuming that the intracluster gas and the dark matter are in hydrostatic equilibrium in the gravitational field created by the total mass of the cluster, and by assuming the knowledge of the hot gas density profile, we have obtained a full description of the observationally relevant properties of BEC dark matter. Then, by using a least-$\chi^2$ fitting between the theoretically derived BEC dark
matter mass and the observed cluster mass, by taking into account the observational data for 106 clusters \cite{RB}, we have obtained the values of the parameter $\lambda =m^3/a$ describing the physical properties of the galactic cluster condensate. Astrophysical observations also provide some numerical constraints on the self-interaction cross section of the dark matter. By combining these two sets of information the mass and the scattering length of the dark matter particle can be constrained.

We have investigated first the zero temperature BEC dark matter model, by considering
that the dark matter temperature is much lower than the transition one, $T\ll
T_{tr}$. In this case, the best fitting result is that $\lambda=\left(1.766-4.755\right)\times 10^{-96} \;{\rm g^3/cm}$. With the upper
limit for self-interaction cross-section, $\sigma_m\leq1.25\;\text{cm$^2\slash$%
g}$, we derive the upper limits for the dark matter particle mass and scattering length as $%
m\leq 1.763\;\mu\text{eV}$, and $a\leq 1.770\times 10^{-7}$ fm, respectively. In the case of the presence of some temperature gradients in the hot intracluster gas, the limits on the mass and scattering length of the dark matter particle, corresponding to $\lambda =\lambda _1$, are $m<2.5\times 10^{-3}$ meV, and $a<2.1\times 10^{-7}$ fm. These bounds are much tighter as compared to the ones obtained from the study of BEC dark matter at the galactic level, which through Eqs.~(\ref{massg}) and (\ref{ag}) provides the upper limits  $m\leq 0.179$ meV and $a\leq 2\times 10^{-6}$ fm, respectively. However, we would like to point out that the galactic estimates are extremely rough, and they are based on qualitative considerations. In order to obtain some better estimates of the mass and scattering length of the dark matter particle from galactic properties, a detailed statistical analysis, also taking into account the effects of the baryonic matter, is required.

The astrophysical parameters of the galactic clusters essentially depend on the cluster central density $\rho _{DM}(0)$. In particular, $\rho _{DM}(0)$ determines the theoretical radius of the cluster, and therefore it can be obtained from the fitting of the theoretical model with the observational data. Hence this property allows the prediction of the cluster mass-central density relation, as well as of the hot gas temperature and cluster central density relation. It is interesting to note that both of these relations can be mathematically described in terms of a simple linear fit, which is statistically significant. Therefore the predicted relations between cluster central density and total mass of the cluster, as well as the predicted relation between gas temperature and $\rho _{DM}(0)$ may allow a direct astrophysical testing of the BEC dark matter model.

 One of the important advances in astronomy in recent years is the observation of large clusters of galaxies, and of the collisions of such clusters. According to the standard $\Lambda $CDM model, when the collision  of large clusters takes place, the cosmic gas present in the clusters slows down, and heats up. These processes can be observed through the  X-ray radiation emitted by the large amounts of hot ($10^7$ K) gas located in the collision area. On the other hand, stars and the cold matter halos of galaxies travel through each other without experiencing any collision. Thus cold dark matter halos continue their motion along the initial paths of the colliding clusters, feeling only the effects of the gravitational interactions.
Important astrophysical information can be obtained from observations of the weak lensing of the background objects, once they are viewed through the colliding clusters.  In this case the deflection angle of the light is proportional to the mass of the gas at the collision point. On the other hand lensing proportional to the sum of the stellar and dark matter masses is seen at the points where  visible galaxies are located.

A particularly interesting case is that of the colliding clusters Abell 520 \cite{lensn}. The lensing observations of the gas-rich core regions of this cluster,  containing no visible galaxies, can only be explained by assuming the presence of a dark matter core, which is coincident with the X-ray gas peak, but not with any stellar luminosity peak. These observations are not consistent with the collisionless nature of the cold dark matter in
the standard model, and various explanations, like collisional deposition of dark matter, filaments along the line of sight direction, distant background cluster, or distant background cluster, have been proposed \cite{lensn}. On the other hand, as suggested in \cite{Will}, if dark matter has a non-zero self-interaction cross-section, as it is the case for Bose-Einstein Condensates,  then the dark matter halos of the individual galaxies in the cluster cores must experience a drag force from the ambient dark matter of the cluster. This drag force does not affect the stellar components of galaxies, and consequently this will lead to a separation between the stellar and dark matter components. It was also suggested in \cite{BoHa07}, that lensing properties of dark mater halos may provide an effective way to distinguish between Bose-Einstein Condensate dark matter and non-interacting dark matter models. The observations of the  A520 cluster give an estimate of $\sigma _m=\sigma /m \approx 3.8 \pm 1.1$ cm$^2$ g$^{-1}$ for the
self-interaction cross section of dark matter, a value which must be contrasted with the upper limit
 $\sigma _m=\sigma /m < 1.25$ cm$^2$/g obtained from the study of the Bullet Cluster \cite{d5}.

 From a thermodynamic point of view a first and fundamental condition that must be satisfied during the cosmological Bose-Einstein Condensation process is the continuity of the dark matter pressure at the transition point. This condition  uniquely fixes the critical transition density $\rho _{tr }$ from the normal dark matter state to the Bose-Einstein condensed state as \cite{Hac}
\begin{eqnarray}
\rho _{tr}&=&3.868\times 10^{-21}\left(\frac{\sigma ^2}{3\times 10^{-6}}\right)\times
\left(\frac{m}{10^{-33}\;{\rm g}}\right)^3\left(\frac{a}{10^{-10}\;{\rm cm}}\right)^{-1}\;{\rm g/cm^3},
\end{eqnarray}
where  $\sigma ^2=\langle \vec{v}^{\;2} \rangle /3c^2$, and $\langle \vec{v}^{\;2} \rangle$ is the average squared velocity of the bosonic  dark matter particle before the BEC phase transition. Therefore from a physical point of view $\sigma $ can be interpreted as the one-dimensional velocity dispersion.

The critical temperature at the moment of the Bose-Einstein condensation is given by Eq.~(\ref{Ttr}), and it can be obtained as
\begin{eqnarray}
T_{tr}&\approx &\frac{2\pi \hbar ^2}{\zeta (3/2)^{2/3}m^{5/3}k_B}\rho _{tr}^{2/3}=
\frac{\left(2\pi \hbar ^2\right) ^{1/3}c^{4/3}}{\zeta (3/2)^{2/3}k_B}\frac{\left(\sigma ^2\right)^{2/3}m^{1/3}}{a^{2/3}},
\end{eqnarray}
 or, equivalently, in the form
\begin{eqnarray}
T_{tr}&\approx &6.57\times10^3\times\left(\frac{m}{10^{-33}\;{\rm g}}\right)^{1/3}\times
\left(\frac{\sigma ^2}{3\times 10^{-6}}\right)^{2/3}\left(\frac{a}{10^{-10}\;{\rm cm}}\right)^{-2/3}\;K.
\end{eqnarray}

By assuming that before the BEC transition the dark matter particles were relativistic, with $\sigma ^2=1/3$, and by considering a dark matter particle mass of the order of $10^{-39}$ g, with a scattering length of the order of $10^{-21}$ cm, we obtain a transition density of the order of $\rho _{tr}\approx 10^{-21}$ g/cm$^3$, which is around $10^8$ times greater than the critical density of the Universe, $\rho _{cr}\approx 10^{-29}$ g/cm$^3$. The corresponding transition temperature is $T_{tr}\approx 1.41\times 10^{15}$ K, which is much bigger than the temperature of the intracluster gas, which is of the order of a few keV, $T_g\approx 10^8$ K. By using these numerical values we can estimate the parameter $\kappa =m^{1/3}/\rho _{tr}^{1/3}a$ as $\kappa \approx 10^{15}$. Hence, taking into account this large value of $\kappa $, from Eq.~(\ref{rhoT}) it follows that the finite temperature effects are negligibly small in the case of the BEC dark matter in galactic clusters. However, the finite temperature model for BEC dark matter is a model worth to further explore. Many physical parameters, like the transition density $\rho_{tr}$ and
the transition temperature of dark matter are known only approximately. Moreover, the description of the dark matter properties after the BEC transition certainly needs taking into account finite temperature effects.

An important issue in our study is to try to determine which of the three considered models provides the best fitting of the observational data. Based on the values of $Z_M^2$ and $Z_R^2$ in Table~\ref{tab:models} one can make some assessments regarding the  model that best fits the observed physical parameters of the considered clusters.

\begin{center}
\begin{table}
\begin{tabular}{|c|c|c|c|}
\hline
Model& $Z_M^2 $ & $Z_R^2$ & Comments\\ \hline
$T_{BEC} = 0$, isothermal & 2232.63 & 1249.77 & $\lambda = 1.76624$, $\rho _{DM}(0) = 1.90934$\\ \hline
$T_{BEC} = 0$, non-isothermal & 2273.02 & 49.5034 & $ \lambda = 4.20896$, $\rho _{DM}(0) = 4.10184$ \\ \hline
$T_{BEC} \neq 0$, isothermal & 2232.63 & - &  $\lambda _0 = 1.76624$ and $\{t,\kappa\}$ as in Fig.~\ref{fig:diffl}\\ \hline
\end{tabular}
\caption{Values of the test functions $Z_M^2$ and $Z_R^2$ for the different models analyzed in this paper. $\lambda$ is given in units of $10^{-96}\rm{g^3/cm}$ and $\rho _{DM}(0)$ in units of $10^{-13}\rm{g/cm^3}$.}\label{tab:models}
\end{table}
\end{center}

 As one can see from the Table~\ref{tab:models}, the $Z_M^2$ test function does not change considerably, and there is only a $2\%$ difference between models. However, the value of the $Z_R^2$ test function is $25$ times bigger for the $T_{BEC} = 0$, isothermal case than for the non-isothermal case. We consider this as {\it a statistical evidence that the non-isothermal intracluster hot gas approach should be the preferred paradigm in the analysis of the astrophysical properties of galactic clusters.}

 In the present paper we have investigated the properties of the Bose-Einstein condensate dark matter at the galactic cluster scale. But dark matter in its condensate form is also present at the galactic level. In order to obtain a consistent description of the BEC dark matter, the physical properties derived from galaxy and galaxy cluster observations must coincide (in the ideal case), or, due to the uncertainties in the astronomical and astrophysical data, show at least an order of magnitude similarity.  The properties of the dark matter particles in the condensate are described by two fundamental parameters, the mass $m$ and the scattering length $a$. From the galactic scale observations the range of $a$ is given by Eq.~(\ref{ag}) as $a<2\times 10^{-6}$ fm, and by Eq.~(\ref{ag1}) as $a\in (5-27)\times 10^{-8}$ fm.  The estimation of $a$ for the dark matter from galaxy cluster observational data gives, via Eq.~(\ref{a1n}), the value $a\approx 2\times 10^{-7}$ fm. By taking into account the precision of the astronomical data, we may state that these values of $a$, obtained from observations at two very different astrophysical scales, indicate at least an order of magnitude agreement with each other. We would also like to point out that the galactic scale values of $a$ are strongly dependent on the (poorly known) radius of the dark matter halo, for which we have adopted a standard value of $R=10$ kpc. Similar correlations can be obtained in the case of the mass of the dark matter particle. The galactic scale values are given by Eqs.~(\ref{massg}) and (\ref{massg1}) as $m<0.2$ meV, and $m\in \left(0.05-0.09\right)$ meV, respectively, while the analysis of the 106 galactic clusters provides for the mass of the dark matter particle a value of the order of $m\approx 1.7\times 10^{-3}$ meV, given by Eq.~(\ref{mass1c}). The differences in the mass values are much larger than those in the numerical values of the scattering lengths. It is matter of further study to find out if these differences are due to the uncertainties in the astronomical/astrophysical data, to the limitations of the used theoretical models, to the statistical methods used in data analysis, or if they show an intrinsic conflict between the BEC dark matter models at the galactic and extragalactic scales.

\section*{Acknowledgement}

We would like to thank to the anonymous referee for comments and suggestions that helped us to improve our manuscript. S.-D. L. gratefully acknowledges financial support for this project from the Fundamental Research Fund of China
for the Central Universities. GM is partially supported by a grant of the Romanian National Authority of Scientific Research, Program for research - Space Technology and Advanced Research - STAR, project number 72/29.11.2013.

\end{document}